\documentclass[11pt]{article}
\pdfoutput=1
\usepackage{draft}
\usepackage{comment}
\usepackage[normalem]{ulem}
\usepackage{hyperref}
\hypersetup{
 colorlinks=true,
 urlcolor=blue,
 anchorcolor=blue,
 citecolor=blue,
 filecolor=blue,
 linkcolor=blue,
 menucolor=blue,
 pagecolor=blue,
 linktocpage=true,
}
\usepackage{graphicx,color,subfig}
\usepackage{cite}
\usepackage{empheq}

\usepackage{bbold}
\usepackage[usenames,dvipsnames,table]{xcolor}
\graphicspath{{./figures/}}
\usepackage{tikz}

\newcommand{\ZZ}{\mathbb{Z}}
\newcommand{\RR}{\mathbb{R}}

\newcommand{\kernel}{\mathbb{K}}	
\newcommand{\modS}{\mathbb{S}}	
\newcommand{\fusion}{\mathbb{F}}	
\newcommand{\id}{\mathbb{1}}	

\newcommand{\op}{\mathcal{O}}
\newcommand\calo{\mathcal{O}}
\newcommand\half{{1\over 2}}

\newcommand{\sbmatrix}[1]{
{\tiny\arraycolsep=0.3\arraycolsep\ensuremath{\begin{bmatrix}#1\end{bmatrix}}}
}

\begin{document}

\unitlength = .8mm

 \begin{titlepage}
 \begin{center}

 \hfill \\
 \hfill \\

\title{Universal Dynamics of Heavy Operators in CFT$_2$}

 \author{Scott Collier,$^{\alpha}$ Alexander Maloney,$^{b}$ Henry Maxfield,$^{P}$ Ioannis Tsiares$^{b}$}

 \address{
 $^{\alpha}$Jefferson Physical Laboratory, Harvard University, 
 Cambridge, MA 02138, USA
 \\
 $^{b}$Department of Physics, McGill University,
 Montreal, QC H3A 2T8, Canada\\
 $^{P}$Department of Physics, University of California,
 Santa Barbara, CA 93106, USA
 }

 \email{scollier@g.harvard.edu, alex.maloney@mcgill.ca, hmaxfield@physics.ucsb.edu, ioannis.tsiares@mail.mcgill.ca}

 \end{center}

 \abstract{
We obtain an asymptotic formula for the average value of the operator product expansion coefficients of any unitary, compact two dimensional CFT with $c>1$.  This formula is valid when one or more of the operators has large dimension or -- in the presence of a twist gap -- has large spin.  Our formula is universal in the sense that it depends only on the central charge and not on any other details of the theory.  This result unifies all previous asymptotic formulas for CFT$_2$ structure constants, including those derived from crossing symmetry of four point functions, modular covariance of torus correlation functions, and higher genus modular invariance.  We determine this formula at finite central charge by deriving crossing kernels for higher genus crossing equations, which give analytic control over the structure constants even in the absence of exact knowledge of the conformal blocks.  The higher genus modular kernels are obtained by sewing together the elementary kernels for four-point crossing and modular transforms of torus one-point functions. Our asymptotic formula is related to the DOZZ formula for the structure constants of Liouville theory, and makes precise the sense in which Liouville theory governs the universal dynamics of heavy operators in any CFT. The large central charge limit provides a link with 3D gravity, where the averaging over heavy states corresponds to a coarse-graining over black hole microstates in holographic theories. Our formula also provides an improved understanding of the Eigenstate Thermalization Hypothesis (ETH) in CFT$_2$, and suggests that ETH can be generalized  to other kinematic regimes  in two dimensional CFTs.
 }

 \vfill

 \end{titlepage}

\eject

\begingroup
\hypersetup{linkcolor=black}

\renewcommand{\baselinestretch}{0.93}\normalsize
\tableofcontents
\renewcommand{\baselinestretch}{1.0}\normalsize

\endgroup

\section{Introduction and discussion}\label{sec:intro}

Two dimensional conformal field theories are among the most important and interesting quantum field theories.  They describe important condensed matter and statistical mechanics systems at criticality and, remarkably, possess an infinite dimensional group of symmetries related to local conformal transformations \cite{Belavin:1984vu}. 
In this paper we will be interested in irrational CFTs with $c>1$ and an infinite number of primary states.  Although these theories are not exactly solvable, they are nevertheless under much greater analytic control than their higher dimensional cousins.  In this paper we will describe a particular example of this fact: the dynamics of heavy (i.e. high dimension) operators is universal in two dimensional CFTs, in the sense that these dynamics are determined only by the central charge and not by any other details of the theory. 

\noindent
The basic dynamical data that defines a CFT$_2$ is a list of primary operators {$\calo_i$}, along with 
\begin{itemize}
\item
Their scaling dimensions $\Delta_i\equiv h_i + {\bar h}_i$ and spins $J_i \equiv h_i - {\bar h}_i$, and
\item
The operator product expansion (OPE) coefficients $C_{ijk}$.
\end{itemize}
These data, along with the central charge $c$, uniquely determine the correlation functions of the theory in flat space as well as on an arbitrary surface.
Ideally one would like to solve the constraints of unitarity and conformal invariance to determine the possible allowed values of the $\{h_i, {\bar h}_i, C_{ijk}\}$, and hence completely classify two dimensional CFTs.  In the absence of such a complete classification, however, we will ask a more modest question: which features of this data are \emph{universal} (i.e. true in any conformal field theory) and which are \emph{theory dependent}?

A simple example of a universal feature is the dimension and spin of the identity operator:\footnote{We restrict our attention in this paper to unitary, compact CFTs, defined to have a discrete spectrum with a unique $\mathfrak{sl}(2)$-invariant ground state.  The same approach will, however, apply more generally with some modest modifications.  We focus on theories with $c_L = c_R = c$ for simplicity, but the modification of our results to theories with $c_L\ne c_R$ is straightforward.}
\begin{equation}\label{iden}
h_{\id} = 0 = {\bar h}_{\id}
\end{equation}
which is the same in every CFT$_2$.
A second and somewhat more subtle universal feature is Cardy's formula for the growth of the high energy density of primary states \cite{Cardy:1986ie}:\footnote{Throughout this paper we use the notation $a\sim b$ to denote that ${a\over b}\to 1$ in the limit of interest. We will also use the notation $a\approx b$ to denote that $a$ and $b$ have the same leading scaling in the limit of interest.}
\begin{equation}\label{Cardy}
\rho(h,{\bar h}) \approx \exp\left\{4\pi \left(\sqrt{{(c-1)\, h \over 24}}+\sqrt{{(c-1)\,\bar h\over 24}}\right) \right\}~~~{\rm when}~ h, {\bar h}\to \infty.
\end{equation}
Equation (\ref{Cardy}) is true in any compact CFT$_2$ with $c>1$, and is universal in the sense that it depends only on the central charge $c$ and not on any other details of the theory.  
In fact, these two universal features (\ref{iden}) and (\ref{Cardy}) are closely related: 
they are ``dual," in the sense that they are related by modular invariance.  Cardy's formula is the statement that the identity operator has dimension zero, albeit interpreted in a dual channel in the computation of the torus partition function.  

Every unitary, compact CFT possesses an additional universal feature: the identity operator will appear in the fusion of any operator with itself.  In terms of the OPE data, this means that
\begin{equation}\label{Cii1}
C_{i i \id}=1
\end{equation}
for any operator $\calo_i$.\footnote{We have chosen a basis of operators such that the two-point function is diagonal and canonically normalized, $\langle\calo_i(0)\calo_j(1)\rangle =\delta_{ij}$.}  This leads to the following natural question: what is the corresponding dual universal feature?  In other words, what universal feature do the three point coefficients obey which plays the same role to equation (\ref{Cii1}) as Cardy's formula (\ref{Cardy}) does to equation (\ref{iden})?

We will answer this question in this paper.  The result is a universal asymptotic formula for the average value of the OPE coefficients:
\begin{equation}\label{av}
\overline{C_{ijk}{}^2}  \sim C_0(h_i,h_j,h_k)C_0(\bar{h}_i,\bar{h}_j,\bar{h}_k)
\end{equation}
where
\begin{equation} \label{C0}
C_0(h_i,h_j,h_k) \equiv
\frac{1}{\sqrt{2}}{\Gamma_b(2Q) \over \Gamma_b(Q)^3} {\prod_{\pm\pm\pm}\Gamma_b\left({Q\over 2} \pm i P_i \pm i P_j\pm i P_k\right)  \over \prod_{a\in\{i,j,k\}} 
\Gamma_b(Q +2iP_a) \Gamma_b(Q -2iP_a)}~.
\end{equation}
In this equation $\prod_{\pm}$ denotes a product of eight terms with all possible sign permutations. Here rather than using the central charge $c$ and dimensions $h$ and $\bar h$ to write our formula, we have used the ``Liouville parameters"
\begin{equation}\label{params}
c = 1+ 6 Q^2 = 1 + 6 (b+b^{-1})^2,~~~ h=\alpha (Q-\alpha), ~~~\alpha = {Q\over 2} + i P~.
\end{equation}
Just as with Cardy's formula, this result is universal in the sense that it is true in any (compact, unitary) CFT, and the only free parameter appearing in this formula is the central charge $c$.

In interpreting this formula, a few comments are in order.
The first is that equation (\ref{av}) is an expression for the average OPE coefficient, with the heavy operator weight(s) averaged over all Virasoro primary operators, which is valid for any finite $c>1$.  In this sense, our result differs from most of the previous results in the literature.
The second is that, although we have only written one formula, equation (\ref{av}) is secretly three different formulas hiding in one.  In particular, this formula is valid in three different regimes, and is derived using three types of crossing symmetry. Equation (\ref{av}) holds:
\begin{itemize}
\item
When two operators are taken to be fixed and the third is taken to be heavy, in which case it follows from the crossing symmetry of four-point functions with pairwise identical external operators.
\item
When one operator is fixed and the other two are heavy, in which case it follows from the modular covariance of torus two-point functions of identical operators.
\item
When all of the operators are taken to be heavy, in which case it follows from modular invariance of the genus two partition function.
\end{itemize}
In each case, the averaging taken in equation (\ref{av}) should be understood as an average over the heavy operator(s), but not over the other operators which are held fixed.\footnote{As we will elaborate on below, ``heavy" in this context means that $h$ and $\bar h$ are much larger than both the central charge and the dimensions of the other operators which are held fixed.  For this reason the three different regimes described above are distinct, and there is a-priori no reason to expect to get the same result in each regime.}
The surprising result is that we obtain exactly the same formula in each case.

Various authors have previously considered the asymptotic behaviour of three point coefficients in each of these three separate limits 
\cite{Chang:2015qfa,Chang:2016ftb,Kraus:2016nwo,Kraus:2017ezw,Kraus:2017kyl,Das:2017vej,Das:2017cnv,Brehm:2018ipf,Hikida:2018khg,Romero-Bermudez:2018dim,Basu:2017kzo,Cardy:2017qhl,Song:2019txa,Pal:2019yhz,Michel:2019vnk,Brehm:2019pcx,Besken:2019bsu}.  The asymptotic formulas which were obtained generally relied on detailed computations of the conformal blocks, and -- while correct -- required assumptions about the behaviour of the blocks in certain kinematic regimes or the simplification of large central charge. Our single asymptotic formula \eqref{C0} unifies all of these previous results, and in the darkness binds them.  Moreover, it holds for any finite value of the central charge $c>1$, and interpolates between all of the previously known results in the literature.

Before describing the details of our derivation, in the remainder of the introduction we will describe the strategy underlying our derivation and comment in more detail on the interpretation of this result.

\subsection{The strategy: bootstrap without the blocks}

In order to illustrate our basic strategy, consider the following simple example where one extracts the asymptotic behaviour of OPE coefficients from crossing symmetry of four point functions. Consider the four point function of an operator $\calo$
\begin{equation}\begin{aligned}
\langle \calo(0) \calo(x) \calo(1) \calo(\infty)\rangle &= \sum_{\calo_s} |C_{\calo\calo \calo_s}|^2 x^{h_s-2h_{\calo}}\bar x^{\bar h_s - 2\bar h_{\calo}}
\cr
&= \sum_{\calo_t} |C_{\calo\calo\calo_t}|^2 (1-x)^{h_t-2h_{\calo}}(1-\bar x)^{\bar h_t-2\bar h_{\calo}}
\end{aligned}\end{equation}
where in first line and second lines we have expanded in a basis of intermediate operators in the $S$-channel and $T$-channel, respectively.  In this simple version of the computation the sums run over all operators in the theory, both primaries and descendants, and we are not organizing the states into representations of the conformal group.  The functions $x^{h_s-2h_{\calo}}$ and $(1-x)^{h_t-2h_{\calo}}$ play the role of conformal blocks in the $S$- and $T$-channel, respectively.
This four point function has a pole at $x=1$ coming from the operator $\id$ in the $T$-channel, which allows us to determine the asymptotic behaviour of the $S$-channel expansion coefficients $|C_{\calo\calo\calo_s}|^2$ when $h_s$ is large. We do so by expanding the $T$-channel conformal block of the identity operator into $S$-channel blocks: 
\begin{equation}\label{binomial}
	\frac{1}{(1-x)^{2h_\calo}} = \sum_{n=0}^\infty (-1)^n \binom{-2h_\calo}{n} x^{n} = \sum_{n=0}^\infty \binom{2h_\calo+n-1}{n} x^n
\end{equation}  
The binomial coefficient $\binom{2h_\calo+n-1}{n}$ appearing in this expression is a simple example of a {\it crossing kernel}: the coefficients which appear when we expand a conformal block in one channel in terms of conformal blocks in a dual channel.\footnote{Note however that this crossing kernel is only supported on a discrete set of intermediate operator weights (namely $h_s = 2h_\calo + n$ for $n$ a non-negative integer); this is similar to the situation for global $SL(2,\RR)$ conformal blocks, which can be expanded as a sum over double-twist blocks and their derivatives in the cross channel (see \cite{Collier:2018exn} for an explicit decomposition).  This is unlike the case of Virasoro blocks that will be the subject of this paper, as the cross-channel decomposition of the Virasoro block will typically involve a continuum.}
Comparing the two channel decompositions of our correlation function, we see that our crossing kernel must equal the average value of the OPE coefficients at $h_s=2h_\op +n$ in the limit where the operator $\calo_s$ is heavy:
\begin{equation}\label{dumbresult}
\overline{|C_{\calo\calo\calo_s}|^2}_\text{scaling} \sim \binom{h_s-1}{h_s-2h_\calo}\binom{\bar{h}_s-1}{\bar{h}_s-2\bar{h}_\calo} \sim \frac{h_s^{2h_\calo-1}}{\Gamma(2h_\calo)} \frac{\bar{h}_s^{2\bar{h}_\calo-1}}{\Gamma(2\bar{h}_\calo)},~h_s,\bar{h}_s\to\infty
\end{equation} 
The subscript `scaling' reminds us that, as we did not organize into representations of the conformal group, the average here is over {all} heavy operators $\calo_s$ -- both primaries and descendants -- of dimensions $h_s,\bar{h}_s$.  We have also not specified the exact nature of the average which is being taken, i.e.\ over how wide a range of operators one must average in order for the result (\ref{dumbresult}) to hold.  We will return to this subtlety below.

In order to determine the asymptotic behaviour of primary operator OPE coefficients we must improve this computation by organizing the sum over intermediate states into a sum over representations of the conformal group.  This is accomplished by taking $\calo_s$ and $\calo_t$ above to be primary operators and replacing the functions $x^{h_s-2h_\calo}$ and 
 $(1-x)^{h_t-2h_\calo}$ by the appropriate conformal blocks.  
We then expand the identity block for the $T$-channel in terms of the $S$-channel blocks for heavy operators, exactly as in \eqref{binomial}.  
The average value of the primary operator OPE coefficients is then given by the analog of the binomial coefficient appearing in this expansion. As conformal blocks for Virasoro symmetry are not known analytically one might think that this computation is impossible.  Remarkably, this is not the case, as Ponsot and Teschner obtained explicit (but complicated) expressions for the crossing kernel of Virasoro blocks for four-point functions \cite{Ponsot:2000mt,Ponsot:1999uf}.\footnote{The higher-dimensional analog of the Virasoro fusion kernel is the $6j$ symbol for the principal series representations of the Euclidean global conformal group $SO(d+1,1)$ \cite{Liu:2018jhs}, which serves as a crossing kernel for conformal partial waves.}
However, when we take the operator in the $T$-channel to be $\id$ these crossing kernels simplify considerably, and they are essentially given by our expression \eqref{C0}.

This computation will be carried out in more detail below, but already several features are apparent.  The first is that, as conformal blocks are purely kinematic objects -- i.e. they depend on central charge and the dimensions of the operators under consideration but not on which theory we are studying -- the crossing kernels are purely kinematic as well.  This guarantees that our resulting asymptotic formula will be universal, in the sense that it depends only on the central charge but not on any other details of the theory.  
The second is that, from this point of view, conformal blocks can be bypassed altogether and one can work directly with crossing kernels.  
In particular, as long as one is interested in understanding the constraints that crossing symmetry imposes on the dynamical data of a CFT (the spectrum and OPE coefficients) the conformal blocks represent an unnecessary complication.  Blocks are only needed if one wishes to extract an observable, such as a correlation function, from this basic dynamical data.

The above discussion shows that crossing symmetry of four point functions will determine the asymptotic behaviour of OPE coefficients in the limit where one operator is taken to be heavy and the others are held fixed.  In order to obtain other constraints, we must consider crossing symmetry and modular invariance for more general observables.  The most general observable is an $n$-point correlation function of Virasoro primaries on a Riemann surface of genus $g$, which we will denote
$G_{g,n}(\{q_i\})$, where the $q_i$ are a set of continuous variables which parameterize the moduli of the Riemann surface as well as  the locations of the insertion points of these primary operators.
We then expand this observable as a sum over intermediate operators propagating in a particular channel, as
\begin{equation}
\begin{aligned}\label{decomp}
	G_{g,n}(\{q_i\}) &= \sum_{\{\mathcal{O}_j\}}\mathcal{C}_{\{\calo_j\}}\mathcal{F}(\{ P_j\}|\{q_i\})\\
	&\equiv \int [dP_j]\,\rho(\{P_j\})\mathcal{F}(\{ P_j\}|\{q_i\}).
\end{aligned}
\end{equation}
Here the $\{\mathcal{O}_j\}$ are the internal operators which contribute to this observable, and the $\mathcal{C}_{\{O_j\}}$ are the corresponding products of OPE coefficients.   We are organizing into conformal families, and the conformal block $\mathcal{F}(\{ P_j\}|\{q_i\})$ encodes the contribution of all descendants of the operators $\{\calo_j\}$. 
As the conformal blocks are kinematic, they depend only on the spins and dimensions of the operators $\{\calo_j\}$, which we are writing in terms of the parameters $\{ P_j\}$ defined by equation \eqref{params}. In order to keep the notation compact, in this formula $\{P_j\}$ and $\{q_i\}$ denote both the holomorphic and anti-holomorphic weights of the internal operators and moduli of the punctured Riemann surface, and the block  $\mathcal{F}(\{ P_j\}|\{q_i\})$ includes contributions from both left- and right-moving descendants.
For simplicity we have suppressed the dependence on the external operators.  In the last line we have introduced a ``density of OPE coefficients'' 
\begin{equation}
\rho(\{P_j\})=\sum_{\{\mathcal{O}_j\}}\mathcal{C}_{\{ O_j\}} \prod_j \delta\left(P_j - P_{\mathcal{O}_j}\right) \delta\left({\bar P}_j - {\bar P}_{\mathcal{O}_j}\right)
\end{equation}
which is a function only of the $P_j$.\footnote{Strictly speaking $\rho$ is a distribution rather than a function.  Moreover, the $P_i$ will be either real or purely imaginary depending on dimensions and spins of the operators $\calo_j$, and the definition of the integral in \eqref{decomp} includes contributions from all states.}

In (\ref{decomp}) we have reduced the correlation function to a sum of products of OPE coefficients.  On a higher genus Riemann surface this is an in principle complicated procedure, as one must decompose the Riemann surface into pairs-of-pants and then sum over internal operators which propagate through the cuffs of these pairs of pants.  This makes the computation of the conformal blocks quite difficult.  The advantage of our approach is that by working directly with crossing kernels rather than conformal blocks, almost all of the details of this construction are irrelevant.  Thus it is possible to understand the constraints of modular invariance and crossing symmetry without the need to explicitly construct the Riemann surface.

We now wish to compare this to the expansion of our observable in another channel:
\begin{equation}\label{decomp2}
\begin{aligned}
	G_{g,n}(\{q_i\}) &= \sum_{\{\mathcal{O}_k\}}\widetilde{\mathcal{C}}_{\{ \calo_k\}}\widetilde{\mathcal{F}}(\{ R_k\}|\{\tilde q_i\})\\
	&= \int [dR_k]\tilde\rho(\{R_k\})\widetilde{\mathcal{F}}(\{ R_k\}|\{\tilde q_i\}).
\end{aligned}
\end{equation}
Here we denote the OPE coefficients, the Virasoro conformal blocks, and the OPE spectral density in this alternate channel with a tilde. 
We have also denoted the moduli on which the conformal blocks depend with a tilde to emphasize that the blocks in different channels typically admit perturbative expansions in different parameterizations of the moduli.  In general the relationship between the two coordinate systems $q_j$ and ${\tilde q}_i$ on moduli space is quite complicated.
Our strategy of working entirely with crossing kernels ensures, however, that we never need to determine this relationship explicitly.

Associativity of the operator product expansion implies that our two different operator product expansions must agree.
We then compare these two different expansions by introducing the crossing kernel $\kernel$ defined by:
\begin{equation}
	\mathcal{F}(\{ P_j\}|\{q_i\}) = \int[d R_k]\kernel_{\{ R_k\}\{ P_j\}}\widetilde{\mathcal{F}}(\{ R_k\}|\{\tilde q_j\}).
\end{equation}
Plugging this into equation ({\ref{decomp}) and comparing with (\ref{decomp2}) gives us the crossing equation.
\begin{equation}\label{crossingK}
\tilde\rho(\{R_k\}) =  \int [dP_j]\, \kernel_{\{ R_k\}\{ P_j\}} \rho(\{P_j\})~.
\end{equation}
In cases where the same OPE data appears in both channels, the solutions to the crossing equation are the unit eigenvectors of the crossing kernel.

We now wish to extract universal features of the OPE coefficients $\mathcal{C}_{\{ O_j\}}$ by considering limits where the identity operator dominates in one channel. 
In particular, we would like to consider cases where the right hand side of the crossing equation (\ref{crossingK}) is dominated by the identity operator (i.e. dominated by the term with all $\calo_j =\id$)  when the internal weights $R_k$ are taken to infinity.  This will occur when
\begin{equation}
	{\kernel_{\{R_k\}\{P_j\}}\over \kernel_{\{R_k\}\{\id\}}}\to 0~~~~~{\rm as}~R_k\to\infty.
\end{equation}
In this limit the density of OPE coefficients is just given by the corresponding crossing kernel of the identity operator:
\begin{equation}
	\tilde\rho(\{R_k\}) \approx \kernel_{\{R_k\}\{\id\}}~~~~~{\rm as}~R_k\to\infty.
\end{equation}
This is the generalization of our earlier result (\ref{dumbresult}), that the crossing kernel of the identity operator serves as the universal asymptotic behaviour of the OPE coefficients for heavy states.

We emphasize that, although we have phrased it more abstractly, this is equivalent to the 
familiar strategy where one studies the crossing equation in a kinematic regime in which the exchange of the identity operator dominates in one channel.  For example, in the case of the four-point function the limit we are considering is equivalent to the one where the cross ratio $x\to1$.  Similarly, the application of this strategy to the torus partition function gives Cardy's formula.
A final example is the lightcone bootstrap \cite{Komargodski:2012ek,Fitzpatrick:2012yx}, where the spectrum and OPE data of  CFT$_{d>2}$ approaches that of mean field theory at large spin.
However these arguments typically require the detailed knowledge of conformal blocks in certain Lorentzian kinematic regimes, which in the Virasoro case is out of reach except in the simplest cases.
The advantage of our approach is that we only require the crossing kernel,  bypassing the need to compute the conformal blocks explicitly.

\subsection{The Moore-Seiberg construction of crossing kernels}

We now wish to apply this construction to constrain the asymptotics of the squared OPE coefficients $|C_{ijk}|^2$.  To begin, recall that $C_{ijk}$ is the correlation function $\langle {\mathcal{O}}_i {\mathcal{O}}_j {\mathcal{O}}_k \rangle_{S^2}$ on the sphere, with the operators inserted at three points.  Thus to study $|C_{ijk}|^2$ we must consider observables obtained by sewing together two copies of the sphere at these insertion points.  For example, the four point function on the sphere is obtained by sewing together these two spheres at a single point -- say, the insertion point of the operators ${\mathcal{O}}_k$ -- to give:\footnote{The notation $\mathcal{O}'(\infty)$ means $\lim_{z\to \infty}z^{2h_{\mathcal{O}}}\bar z^{2\bar h_{\mathcal{O}}}\mathcal{O}(z,\bar z)$.}
\begin{equation}
\langle {\mathcal{O}}_i(0) {\mathcal{O}}_j(x,\bar x) {\mathcal{O}}_j(1) {\mathcal{O}}'_i(\infty) \rangle_{S^2} = \sum_{\mathcal{O}_k} |C_{ijk}|^2 \mathcal{F}({P_k}|x)\overline{\mathcal{F}}(\bar P_k|\bar x)
\end{equation}
where $\mathcal{F}({P_k}|x)$ is an appropriate holomorphic conformal block.
Applying the crossing arguments of the previous section will then lead to an asymptotic formula for the $|C_{ijk}|^2$ in the limit where ${\mathcal{O}}_k$ is taken to be heavy but the operators ${\mathcal{O}}_i$ and ${\mathcal{O}}_j$ are held fixed.  Similarly, we can sew together the spheres at a pair of points, the locations of the operators ${\mathcal{O}}_j$ and ${\mathcal{O}}_k$, to obtain the two point function on the torus:
\begin{equation}\label{tortwo}
\langle {\mathcal{O}}_i(v,\bar v){\mathcal{O}}_i(0) \rangle_{T^2} (\tau) = \sum_{\mathcal{O}_j, \mathcal{O}_k} |C_{ijk}|^2 \mathcal{F}({P_j, P_k}|\tau,v)\overline{\mathcal{F}}(\bar P_j,\bar P_k|\bar\tau,\bar v)
\end{equation}
where $\mathcal{F}({P_j,P_k}|\tau,v)$ is now a conformal block for two point functions on the torus.
This will lead to an asymptotic formula for $|C_{ijk}|^2$ in the limit where both $ {\mathcal{O}}_j$ and $\mathcal{O}_k$ are heavy and $\mathcal{O}_i$ is fixed.  Finally, sewing together all three insertion points gives the genus two partition function:
\begin{equation}\label{gtwo}
Z_{g=2}(q,\bar q)= \sum_{\mathcal{O}_i, \mathcal{O}_j, \mathcal{O}_k} |C_{ijk}|^2 \mathcal{F}({P_i, P_j, P_k}|q)\overline{\mathcal{F}}(\bar P_i,\bar P_j,\bar P_k|\bar q)
\end{equation}
where $q$ is a collection of genus two modular parameters and 
$\mathcal{F}({P_i, P_j,P_k}|q)$ is a holomorphic genus two conformal block.  
This will lead to an asymptotic formula which is valid when all of the operators are taken to be heavy.

The strategy described above is only useful, however, if we can accomplish two things: we first need to find a dual channel where the identity operator dominates, and we must then compute the relevant crossing kernels.  To accomplish this we will follow the strategy of Moore and Seiberg \cite{Moore:1988qv}, who argued that all of the constraints of the associativity of the OPE are completely captured by crossing symmetry of four point functions on the sphere and modular covariance of one-point functions on the torus.  This is because any crossing transformation for any observable can be constructed by composing ``elementary" crossing transformations: four point crossing on the sphere (or fusion), and modular transformations for one-point functions on the torus (along with braiding, which we will not use in this paper).
The crossing kernels for these elementary crossing transformations were written down explicitly in \cite{Ponsot:1999uf,Ponsot:2000mt,Teschner:2001rv,Teschner:2003at}.  Thus, by assembling these together using the Moore-Seiberg construction, we can obtain explicit formulas for general crossing transformations -- such as those on higher genus Riemann surface -- without ever computing a conformal block.

We will write this down very explicitly below, but the general strategy is easy to understand.  The two elementary crossing transformations we use can be represented pictorially as in figure \ref{fig:G11ex}.
\begin{figure}
	\centering
	\includegraphics[width=.125\textwidth]{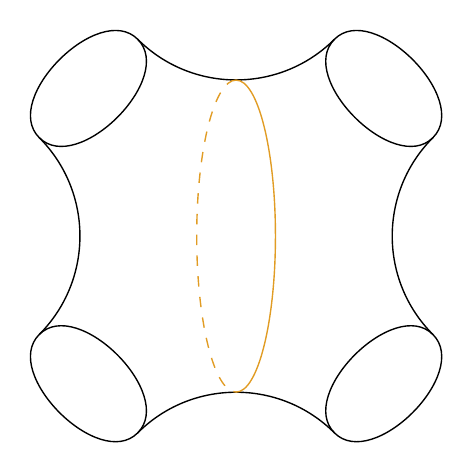} \raisebox{25pt}{\scalebox{1}{$\quad\xrightarrow{\quad\quad}\quad$}} \includegraphics[width=.125\textwidth]{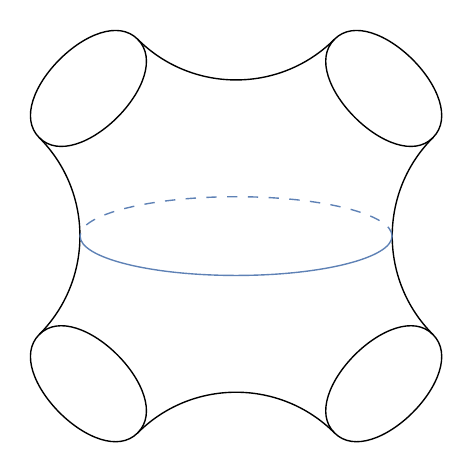}  
\raisebox{6pt}{\raisebox{20pt}{~~~~~and~~~~~~}	
		\includegraphics[width=.15\textwidth]{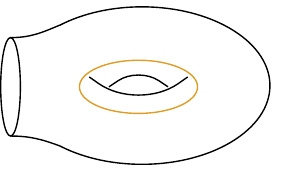} \raisebox{20pt}{\scalebox{1}{$\quad\xrightarrow{\quad\quad}\quad$}} \includegraphics[width=.15\textwidth]{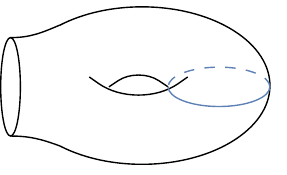}}
\caption[The elementary crossing transformations.]{The elementary crossing transformations: sphere four-point crossing between $S$ and $T$ channels, and torus one-point crossing between the $\tau$ and $(-1/\tau)$ frames.
	\label{fig:G11ex}}
	\end{figure}
The first of these is the crossing transformation for four point functions on the sphere, where we have chosen to represent the four external operators by holes rather than infinitesimal points.
The $S$- and $T$-channel decompositions of the four point function then correspond to the two different ways of constructing this four-holed sphere as two pairs-of-pants glued together shown above.  Similarly, the second picture in figure \ref{fig:G11ex} describes the crossing transformation between two different channels for a one-point function on the torus.

We can now construct crossing transformations for two point functions on the torus by composing these elementary transformations, as in figure \ref{fig:G12ex}.
\begin{figure}
	\centering
	\includegraphics[width=.2\textwidth]{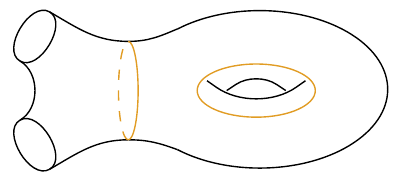} \raisebox{.034\textwidth}{\scalebox{1.5}{$\quad\xrightarrow{\quad\quad}\quad$}} \includegraphics[width=.2\textwidth]{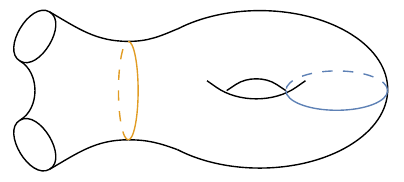}\raisebox{.034\textwidth}{\scalebox{1.5}{$\quad\xrightarrow{\quad\quad}\quad$}} \includegraphics[width=.2\textwidth]{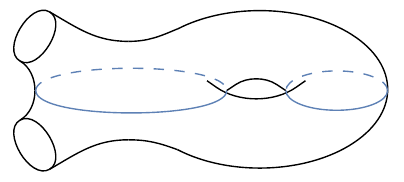}
\caption[Example 1.]{Example of a crossing transformation on the torus two-point function. 
	\label{fig:G12ex}}
	\end{figure}
We recognize the first of these as the modular S transformation for one point functions on the torus, and the second as the fusion move for four point functions on the sphere.  The result is an expression for this more complicated crossing kernel as a product of these two elementary kernels.  Indeed, we recognize the channel on the far right as precisely the one which gives the square of the OPE coefficients in equation (\ref{tortwo}), where $\mathcal{O}_j$ and $\mathcal{O}_k$ are the operators which propagate through the two blue circles.  Our asymptotic formula for $|C_{ijk}|^2$ is then obtained by considering the kinematic limit which is dominated by the identity operator $\id$ propagating in the channels (marked by yellow circles) on the far left.

We can construct the crossing transformations at genus two in a similar manner, as in figure \ref{fig:G13ex}:
\begin{figure}
	\centering
	\includegraphics[width=.215\textwidth]{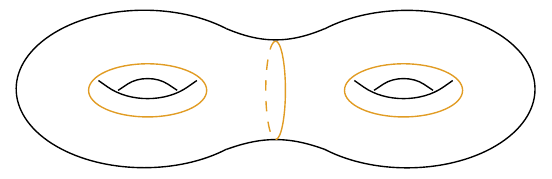} \raisebox{.034\textwidth}{\scalebox{1.5}{$\quad\xrightarrow{\quad\quad}\quad$}} \includegraphics[width=.215\textwidth]{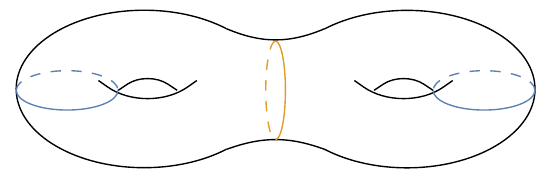}\raisebox{.034\textwidth}{\scalebox{1.5}{$\quad\xrightarrow{\quad\quad}\quad$}} \includegraphics[width=.215\textwidth]{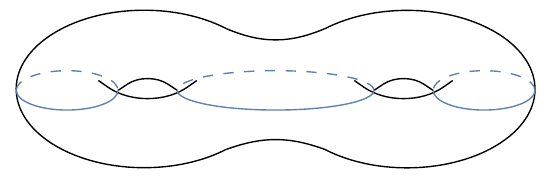}
\caption[Example 2.]{Example of a crossing transformation on the $g=2$ partition function. 
	\label{fig:G13ex}}
	\end{figure}
we have first done two crossing moves for torus one point functions, followed by a four-point crossing move on the sphere.  
Again, the channel on the far right gives the square of the OPE coefficients considered in equation (\ref{gtwo}) where the operators $\calo_i$, $\calo_j$ and $\calo_k$ propagate through the three blue circles.  The asymptotic formula for $|C_{ijk}|^2$ when these three operators are taken to be heavy is found by considering the limit where the identity operator $\id$ dominates in the channel decomposition depicted on the far left.  This formula is given in terms of a genus two crossing kernel which -- by construction -- is a product of the elementary crossing kernels which were written down by Ponsot and Teschner.

The result is an asymptotic formula for the averaged OPE coefficients $|C_{ijk}|^2$ in the three limits described above, where either one, two or all three operators are taken to be heavy, and only the heavy operators are averaged over. For example, in the case where the differences between the heavy operator dimensions and all spins $J_i$ are held fixed in the large-dimension limit, we can state all of our asymptotic formulas as follows:\footnote{Here, our notation with the $\approx$ symbol means that we have omitted the order one coefficients appearing in these formulas. These coefficients can be found in equations \eqref{eq:C0HLL}, \eqref{eq:C0HHLfixed} and \eqref{eq:C0HHHfixed}.}
\begin{align}
	\overline{C^2_{\calo_1\calo_2\calo}} &\approx 16^{- \Delta}e^{-2\pi\sqrt{{c-1\over 12}\Delta}}\Delta^{2(\Delta_1+\Delta_2)-{c+1\over 4}},~&&\Delta\gg c,J,\Delta_i,J_i\label{eq:HLL}\\
	\overline{C^2_{\calo_0\calo_1\calo_2}} & \approx e^{-4\pi\sqrt{{c-1\over 12}\Delta_1}}\Delta_1^{\Delta_0},~&&\Delta_1,\Delta_2\gg c,J_i,\Delta_0,J_0,|\Delta_1-\Delta_2|\label{eq:HHL}\\
	\overline{C^2_{\calo_1\calo_2\calo_3}} & \approx \left({27\over 16}\right)^{3\Delta_1}e^{-6\pi\sqrt{{c-1\over 12}\Delta_1}}\Delta_1^{5c-11\over 36},~&&\Delta_1,\Delta_2,\Delta_3\gg c,J_i,|\Delta_i-\Delta_j|\label{eq:HHH}
\end{align}
In addition to these, there are other distinct asymptotic limits, for example fixing the ratios of $\Delta_i$ instead of differences as in \eqref{eq:C0HHL} and \eqref{eq:C0HHH}, which are also controlled by \eqref{av}.
Remarkably, all of these formulas (appearing in equations \eqref{eq:C0HLL}, \eqref{eq:C0HHL}, \eqref{eq:C0HHLfixed}, \eqref{eq:C0HHH} and \eqref{eq:C0HHHfixed}) are realized as limits of the same underlying formula \eqref{C0}.
This is perhaps the most surprising feature of our result, and is a consequence of the Moore-Seiberg procedure which constructs all of these different crossing kernels from the same elementary building blocks.  

\subsection{Generalizations to other observables}

We emphasize that, although we have applied our strategy to the computation of the asymptotics of the $|C_{ijk}|^2$, this argument works much more generally.  Whenever one can find a kinematic limit where the identity block dominates a CFT observable, there is a corresponding universal formula for the OPE data in the dual channel -- it is just a matter of assembling the appropriate crossing kernel.  In this sense our strategy should be regarded as defining an entire class of CFT asymptotic formulas which govern the universal dynamics of heavy operators in two dimensional CFTs. It would clearly be worth exploring these dynamics in more detail.

In addition, while our main focus is on {\it universal} asymptotic formulas -- namely those which are constructed only from the propagation of the identity operator in a cross channel -- one can also consider non-universal quantities which are constructed from other light operators propagating in a cross channel.  For example, the leading corrections to the universal formulas described above will come from the other light operators in the theory, and one can obtain improved (but non-universal) asymptotic formulas which depend on the data (such as the spectrum and OPE coefficients) of whatever light operators are present in the theory.

The most interesting example of this type would be one where the contribution from $\id$ in the cross channel vanishes, in which case the asymptotic behaviour would be non-universal and depend on the light data of the theory.
The prototypical example is the average value of the Light-Heavy-Heavy OPE coefficient $\overline{C_{iij}}$, where the state $i$ is heavy and averaged over, while the $j$ is held fixed. This is determined by considering the modular covariance of one point functions $\langle \calo_j\rangle_{T^2}(\tau)$ on the torus in the limit $\tau\to0$ \cite{Kraus:2016nwo}.  The contribution from the identity operator propagating in the dual channel (i.e. taking $\tau\to-1/\tau$) is just the one-point function of $\calo_j$ on the plane, which vanishes.    The first non-vanishing contribution will come from the lightest operator $\chi$ which has $C_{j\chi\chi}\ne0$.  Previous results have either worked only at large central charge, or have organized into scaling blocks or global blocks, rather than full conformal blocks (so that the average in $\overline{C_{iij}}$ is an average over quasi-primaries or over all states in the theory, rather than over Virasoro primaries) \cite{Kraus:2016nwo}.  We can now write down the complete answer at finite central charge, where the average is taken only over primaries; this will be discussed in Section \ref{sec:ETH}.

\subsection{Large central charge limit}

One important special case is the large central charge limit, which is relevant for holographic theories with an AdS gravity dual.  In this case a generic heavy state is interpreted as a microstate of a BTZ black hole.  The observation that the average OPE coefficients take a universal form then has a natural physical interpretation, as the emergence of a semi-classical black hole geometry which arises upon coarse-graining over heavy states.  That our formula depends only on the central charge and the dimensions and spins of the operators reflects the fact that this semi-classical configuration is purely geometric: the holographically computed OPE coefficient depends on Newton's constant and the masses and spins of the objects under consideration, but not on any other details of the state. Our formulas can thus be regarded as an extrapolation of the usual gravitational ``no hair" theorems to CFT.
Indeed, various limits of our formula have already been shown to reproduce the classical dynamics of particles in black hole backgrounds \cite{Kraus:2016nwo,Chang:2016ftb,Brehm:2018ipf, Kraus:2017ezw, Kraus:2017kyl}, and appear in closely related gravitational computations of semiclassical conformal blocks \cite{Maloney:2016kee,Dong:2018esp}. We note that from the point of view of classical gravity it is not at all obvious that there should be a single formula that interpolates between the three different limits we are considering (where either one, two or all three of the operators are taken to be heavy).  Indeed, our formula reflects this: it smoothly interpolates between these three limits at finite $c$, but not after taking a $c\to\infty$ limit.
 
Perhaps the most important point to emphasize here is that, as we take $c\to\infty$, the ``heavy" operators appearing in our formula should still be understood to have dimension large compared to $c$.   This is necessary in order for the identity operator to still dominate in the dual channel.  Such a state, however, will be interpreted as a black hole whose horizon area is very large in AdS units.  A black hole whose size is order one in AdS units would correspond to an operator whose dimension is order $c$.  It is therefore natural to ask under what circumstances the regime of validity of our asymptotic formulas could be extended to operators with finite $h/c$ in the large $c$ limit.  Generically, this will only happen if we impose severe restrictions on the ``light" data in our theory.  For example, the regime of validity of Cardy's formula can be extended all the way down to dimensions of order $c$ only if the density of states of the light spectrum is sufficiently sparse \cite{Hartman:2014oaa}.  It would be interesting to ask whether similar considerations could be applied to our asymptotic formulas.  
We expect that the corresponding sparseness constraint will be considerably more subtle, however, and may require more than just a constraint on the density of OPE coefficients of light operators -- see \cite{Belin:2017nze, Dong:2018esp} for discussions of this in the context of higher genus partition functions of symmetric product orbifolds and holographic CFTs.

\subsection{Chaos, integrability and eigenstate thermalization}\label{sec:introETH}

Our results have an important role to the play in the study of chaos in two dimensional CFTs.  To see this, we first note that while we have written formulas of the form
\begin{equation}\label{av2}
\overline{C_{ijk}{}^2}  \sim C_0(h_i,h_j,h_k)C_0(\bar h_i,\bar h_j,\bar h_k)
\end{equation}
we have not yet stated precisely what range of states one must average over.  The weakest possible statement would be that our asymptotic formula is true only in an integrated sense, where rather than averaging over a small window of states one simply sums over all states below some (large) cutoff.  We expect, however, that a much stronger version is true, where one needs to integrate only over a small window; results that establish this kind of behaviour go under the general name of Tauberian theorems (see e.g.\ \cite{Qiao:2017xif,Mukhametzhanov:2018zja,Mukhametzhanov:2019pzy,Ganguly:2019ksp,Pal:2019yhz,Pal:2019zzr} for recent applications of Tauberian theorems in this context).  In the present case we would require new results for several variables, adapted to the Virasoro crossing transforms.  This is an important avenue for future research, which is not merely a mathematical subtlety but a question of important physical interest.

In particular, our expectation is that in a generic, chaotic theory one would need to average only over very small window in order to obtain the asymptotic result (\ref{av2}).  In other words, in a chaotic theory the typical OPE coefficient should be rather close to the average one.  In an integrable theory, however, many OPE coefficients will vanish due to selection rules, so any average result is obtained only by including many different states in the average.  We expect that in a chaotic theory one would need to average over a window of size not much larger than $e^{-S}$, where $S$ is the microcanonical entropy, while in an integrable theory one must average over a window of some fixed width rather than one that is exponentially small at high energies.  It is important to emphasize that all of our results are derived from crossing and modular constraints which hold in any CFT.  Thus our result \eqref{av2} will be equally true in integrable and chaotic theories.  The crucial difference will be in the {\it way} in which this average is realized.  Indeed, we would propose that the size of the window one must average over should be used as a sharp criterion for chaos in conformal field theory: a chaotic theory is one where one needs to average only over windows of size ${\calo(e^{-S})}$.  It would be interesting to compare this to other proposed characterizations of chaos in quantum field theory.

Indeed, our asymptotic formulas also play an important role in the Eigenstate Thermalization Hypothesis (ETH) \cite{Srednicki:1994ne,deutsch1991quantum}, which states that in a chaotic theory the matrix elements of an operator $\calo$ should obey
\begin{equation}\label{eth}
\langle i | \calo |j \rangle \approx f^{\calo}(\Delta_i) \delta_{ij} + g^{\calo}(\Delta_i, \Delta_j) R_{ij} 
\end{equation}
for states $i$ and $j$ of fixed energy density in a large volume thermodynamic limit.  Here, $f^\calo$ and $g^\calo$ are smooth functions of energy related to the microcanonical one- and two-point functions, and $R_{ij}$ is a random variable of zero mean and unit variance; if the one- and two-point functions are of order one, then $f^\calo$ is of order one and $g^\calo$ of order $e^{-S/2}$. In a scale-invariant theory, the large volume thermodynamic limit is equivalent to a large energy limit at fixed volume, which is the heavy limit we have been studying. When $\calo$ is a local operator, ETH is a statement about the statistics of structure constants (see 
\cite{Asplund:2015eha,deBoer:2016bov,Lashkari:2016vgj,He:2017vyf,Lashkari:2017hwq,Faulkner:2017hll,Guo:2018pvi,Maloney:2018hdg,Maloney:2018yrz,Dymarsky:2018lhf,Dymarsky:2018iwx,Anous:2019yku,Dymarsky:2019etq,Datta:2019jeo,Besken:2019bsu} 
for more detailed discussion of ETH in the context of conformal field theories).  

In a two dimensional CFT it is natural to take this to be a statement about primary operator OPE coefficients; descendant state OPE coefficients are completely determined by Ward identities, and hence by definition do not provide any information about the chaotic dynamics of the theory.  Indeed, dynamics within a particular Virasoro representation will never thermalize due to the infinitude of conserved quantities.  At infinite central charge this distinction is largely irrelevant, as the typical high energy state is -- if not a primary state itself -- then very close to one.  For finite $c$ CFTs, however, these considerations become important and the most sensible definition of ETH is one where (\ref{eth}) is interpreted as a statement about the statistics of primary operators.

In this case our asymptotic formulas for $\overline{C_{\calo ii}}$ and $\overline{C_{\calo ij}{}^2}$ determine the functions $f^\calo$ and $g^\calo$:
 \begin{equation}\label{ETH}
 \overline{C_{\calo ii}} = f^{\calo}(\Delta_i) ,\qquad \overline{|C_{\calo ij}|^2} =  (g^\calo(\Delta_i,\Delta_j))^2
 \end{equation}
 Thus our formulas provide a precise formulation of ETH for CFTs with finite central charge $c$.
  It is important to emphasize that our asymptotic formulas predict the form of the smooth functions $f^\calo$ and $g^\calo$ (and provide the consistency check that $\overline{|C_{\calo ij}|^2}\sim e^{-S}$), but say nothing about the statistics of the remainder term $R_{ij}$.  The statement that $R_{ij}$ has zero mean and unit variance, severely constraining the fluctuations of matrix elements, is an important component of ETH and one which is invisible using the techniques of this paper. Indeed, all CFTs 
are crossing invariant, so no argument based on crossing symmetry alone can distinguish between a chaotic and an integrable theory. Our arguments establish the universal behaviour of averaged OPE asymptotics, and so are not sensitive to the fine-grained statistics of individual eigenstates. 
Some additional input must be included in order to use crossing arguments to probe this more refined structure of ETH. One might hope that assuming no additional currents would be sufficient to ensure the theory is chaotic, but while we make use of this assumption to establish universal formulas that apply at large spin, it is not clear how to use it to say more about statistics of OPE coefficients relevant for ETH.

An important feature of the ETH formula is that it is expected to govern the statistics of OPE coefficients in the Heavy-Heavy-Light limit, where the operators $i$ and $j$ are heavy but $\calo$ is fixed.  On the other hand, our asymptotic formulas for OPE coefficients smoothly interpolate between this limit and the Light-Light-Heavy and Heavy-Heavy-Heavy regimes.  This immediately suggests that the ETH conjecture (\ref{eth}) should be generalized to these regimes as well.  It also suggests that a version of ETH should hold not just at large dimension, but also for operators with large spin at fixed twist.   
We expect this extended regime of validity to be a special feature of CFTs (where there is a state-operator correspondence) rather than general QFTs.  One intriguing aspect of this conjecture is that while the Heavy-Heavy-Light version of ETH has a natural thermodynamic interpretation -- it captures the intuitive notion that in a chaotic theory every state should be approximately thermal in the thermodynamic limit -- the interpretation of equation \eqref{eth} in this extended regime is much more mysterious.

A second important point is that the behaviour of the two functions $f^\calo$ and $g^\calo$ is quite different in two dimensional CFTs from their behaviour in higher dimensions.  In a higher dimensional theory the diagonal terms in the OPE coefficients are exponentially larger than the off-diagonal terms: $f^\calo$ is of order one, while $g^{\calo} \approx e^{-\half S({\Delta_i+\Delta_j\over 2})}$ is exponentially suppressed.  In a two dimensional CFT this behaviour is modified, as $f^\calo$ itself is exponentially small. This can be seen by noting that at high temperature a thermal one point function becomes a one point function on the cylinder $S^1\times \mathbb{R}$, which is -- by the usual radial quantization map -- conformally equivalent to the plane.  Hence thermal one point functions will be exponentially small at high temperature, with exponent determined by the dimension of the lightest operator which couples to the operator $\calo$.  Thus we expect that the off-diagonal terms for a generic primary operator $\calo$ will be exponentially suppressed relative to the diagonal terms, but with an exponent that is not $ e^{-\half S({\Delta_i+\Delta_j\over 2})}$ but rather is determined by the size of the gap in the theory.  This is a consequence of the strange fact that in CFT$_2$ thermal one point functions vanish at high temperature, while thermal two point functions do not.

In the extreme case -- where the size of the gap in the theory is sufficiently large -- the off-diagonal terms will be the same size 
as the diagonal terms. We will clarify this statement in section \ref{sec:ETH} and show that this will occur when the lightest non-vacuum primary that couples to $\mathcal{O}$ has dimension greater than or equal to $c-1\over 16$ (in the case that this lightest operator is a scalar). This fact will be a simple consequence of the structure of the corresponding crossing kernels. A theory with a gap of size ${\cal O}(c)$ would be interpreted as a theory of pure gravity in AdS$_3$ in the large $c$ limit, as the spectrum of perturbations around empty AdS would include only boundary gravitons (i.e. descendants of the identity operator).  We therefore come to a remarkable conclusion -- a theory of pure gravity in AdS$_3$ is precisely one where the off-diagonal terms in ETH are not suppressed relative to the diagonal ones.  This provides an intriguing link between black hole dynamics and quantum chaos.  A similar conclusion was recently reached for JT gravity in two dimensions in \cite{Saad:2019pqd}.

\subsection{Connection to Liouville theory}\label{sec:LiouvilleUniqueness}

Our universal OPE coefficient formula \eqref{C0} closely resembles the DOZZ formula for the structure constants of Liouville theory \cite{Dorn:1994xn,Zamolodchikov:1995aa}. However, our universal asymptotic formulas do not apply to Liouville theory, since it is not compact (the spectrum does not include an $\mathfrak{sl}(2)$-invariant ground state). We here explain the similarity of the formulas by noting that they both follow from Virasoro representation theory, and contrast their interpretation.

The spectrum of Virasoro primary states of Liouville theory is continuous, consisting of scalars of dimension $h=\bar{h}=\frac{c-1}{24}+P^2$ for $P>0$. Their three-point coefficients are given by the DOZZ formula $C_\mathrm{DOZZ}(P_1,P_2,P_3)$, which is related to our formula \eqref{C0} by
\begin{equation}\label{DOZZ}
	C_0(P_1,P_2,P_3) \propto \frac{C_\mathrm{DOZZ}(P_1,P_2,P_3)} {\left(\prod_{k=1}^3S_0(P_k)\rho_0(P_k)\right)^{1\over 2}},
\end{equation}
with a proportionality constant independent of $P_{1,2,3}$, and $S_0$ is the `reflection coefficient' defining the normalisation of the vertex operators through the two-point function\footnote{The proportionality constant is \begin{equation}
	\frac{ (\pi\mu\gamma(b^2)b^{2-2b^2})^{Q\over 2b}}{ 2^{3\over 4}\pi}\frac{\Gamma_b(2Q)}{\Gamma_b(Q)}\end{equation} and the reflection coefficient is
	\begin{equation}
		S_0(P) = (\pi\mu\gamma(b^2)b^{2-2b^2})^{-2iP/b}\frac{\Gamma_b(2iP)\Gamma_b(Q-2iP)}{\Gamma_b(Q+2iP)\Gamma_b(-2iP)},
	\end{equation} where $\mu$ is the Liouville cosmological constant and $\gamma(x) = {\Gamma(x)\over \Gamma(1-x)}$.}
\begin{equation}\label{eq:reflection}
	\langle V_{P_1}(0)V_{P_2}(1)\rangle = 2\pi \delta(P_1-P_2) S_0(P_1).
\end{equation}
Since the theory is noncompact, there is in fact no canonical normalisation of operators, and only the combination \eqref{DOZZ} (up to the $P$-independent normalisation) is unambiguously determined from the bootstrap. The denominator can be understood as a change of measure on the space of states, from the one defined by \eqref{eq:reflection} to a natural one proportional to $dP\, \rho_0(P)$ (see footnote \ref{PlancherelFootnote}).

Given this relation, on might be tempted to interpret our result as describing the precise sense in which Liouville theory captures the universal dynamics of heavy operators, a point of view that has been advocated in the context of holographic theories in \cite{Jackson:2014nla,Turiaci:2016cvo}. We should not, however, interpret this too literally, since $C_\text{DOZZ}$ has a very different interpretation to $C_0$. In particular, Liouville theory has only scalar primary operators, with OPE coefficients $C_\text{DOZZ}$, whereas our results give OPE coefficients for all spins, from a product of two copies of $C_0$ (left- and right-moving). Indeed, a unitary compact CFT with $c>1$ will necessarily contain primary operators with arbitrarily large spin \cite{Collier:2017shs}, and Liouville theory falls outside the scope of our asymptotic formula precisely because it is not compact. Rather, we regard the relation \eqref{DOZZ} as a consequence of the fact that Liouville dynamics is governed by precisely the same Virasoro representation theory that determines our asymptotic formula, as we now explain.

Liouville theory is distinguished by having only scalar Virasoro primary states. In this sense, it is analogous to the A-series or diagonal minimal models which exist for degenerate values of $c<1$, and have a spectrum of scalar primaries (finitely many in that case). The restriction to scalars is sufficient to uniquely specify the theory, since it determines a unique solution to the bootstrap (up to normalisation of operators and a decoupled TQFT). Furthermore, this solution is given explicitly in terms of the identity fusion kernel by a relation precisely of the form \eqref{DOZZ}, which is determined by representation theoretic considerations. We give an argument that can be applied both to four-point crossing symmetry and to modular covariance of torus one-point functions. This type of argument for four-point crossing is not new (see \cite{Ribault:2014hia}, for example), but the version for torus one-point functions is novel, as far as we are aware.\footnote{We thank S.\ Ribault for correspondence.} We sketch the arguments here, giving more detailed explanations of the relevant identities in section \ref{sec:LiouvilleEquations}.

To outline the argument for uniqueness, we first write the crossing equation \eqref{crossingK} including left- and right-moving dependence explicitly as
\begin{equation}
\rho'(P',\bar{P}') =  \int [dP d\bar{P}]\, \kernel_{P'P} \kernel_{\bar{P}'\bar{P}} \,\rho(P,\bar{P}) \,.
\end{equation}
Here, the densities $\rho,\rho'$ denotes a spectral density for internal operators in either the four-point function or the torus one-point function, and $\kernel$ is either a fusion kernel $\fusion$ or a modular S-transform $\modS$, as discussed in sections \ref{sec:fusion} and \ref{sec:modularS} respectively. We can schematically write this as a matrix equation
\begin{equation}\label{eq:matrixCrossingEquation}
	\rho' = \kernel \rho \kernel^\dag,
\end{equation}
where the rows and columns of $\rho$ are labelled by $P,\bar{P}$ respectively, and similarly for $\rho'$. Now, if we assume that the spectrum contains only scalars, then $\rho$ and $\rho'$ are diagonal (nonzero only for $P=\bar{P}$). In that case, we can choose to use a different normalisation for the conformal blocks, and hence fusion kernel, that absorbs factors of $\rho^{1/2}$, $(\rho')^{-1/2}$ into the columns and rows of $\kernel$: $\hat{\kernel} = (\rho')^{-1/2}\kernel \rho^{1/2}$. With this normalisation, the crossing equation becomes $\hat{\kernel} \hat{\kernel}^\dag = \id$, so that $\hat{\kernel}$ is unitary (after restricting to the support of $\rho,\rho'$). Such a normalisation exists for the fusion kernel \cite{Ponsot:1999uf}, thus determining a scalar solution of crossing. This solution reproduces the DOZZ formula up to the ambiguities of normalisation. Moreover, the only way that this solution can fail to be unique is if $\hat{\kernel}$ is block diagonal in the $P$ basis.\footnote{In fact, the Virasoro fusion kernel is block diagonal, since the degenerate representations form an invariant subspace. If we relax the assumption of unitarity, this leads to a second solution to crossing, namely the `generalized minimal model' \cite{Ribault:2014hia}.}

For the final step, we must relate the unitary normalisation of $\kernel$ to the identity fusion kernel. For four-point crossing, such a relation follows from a special case of the pentagon identity satisfied by the fusion kernel. The identity representation is picked out by its simple fusion rule, which implies that the fusion kernel with an external identity operator is trivial. For the torus one-point function, we have a similar identity relating the modular S-matrix and fusion kernel. We give the explicit forms of these identities and their derivations in section \ref{sec:LiouvilleEquations}, along with arguments explicitly verifying them from the closed-form expressions \cite{Ponsot:1999uf,Ponsot:2000mt,Teschner:2003at} for the Virasoro fusion and modular kernels.

 \subsection{Discussion}
 
Before moving on to a derivation of our formula, we discuss a few final interesting features of our result.  

While our asymptotic formula (\ref{C0}) might look arbitrary, it is in fact extremely highly constrained if we assume analyticity.  In fact, equation (\ref{C0}) is almost completely determined by its analytic structure and simple physical considerations.  To see this, we note that $C_0(P_i,P_j,P_k)$ is a meromorphic function of its arguments which has 
\begin{itemize}
\item
Zeroes when $P_i= i {Q\over 2} \pm {i\over 2} \left(r b + s b^{-1} \right)$ with $r,s\in\ZZ_{\ge0}$,
\item
Poles when $P_i = P_j + P_k \pm i{Q\over 2} + i \left(m b + n b^{-1}\right)$ with $m,n\in\ZZ_{\ge0}$,
\end{itemize}
and is invariant under reflections $P_i \to - P_i$ and permutations of the $(P_i, P_j, P_k)$.  
These zeros occur precisely when $\calo_i$ has has a null Virasoro descendant at level $rs$.  The poles occur precisely when the weights of $\calo_i$ are equal to the weights of a double twist operator built out of $\calo_j$ and $\calo_k$ \cite{Collier:2018exn}. 
A meromorphic function is uniquely determined by its poles and zeroes, up to the exponential of a polynomial.  Thus in retrospect, once one postulates the existence of a meromorphic function that interpolates between the asymptotic regimes, one could have completely determined $C_0(P_i, P_j, P_k)$ up to the exponential of a polynomial in the $(P_i, P_j, P_k)$, simply by demanding the existence of zeros at null states and poles at double twist operators. One might even argue that this polynomial must be a constant in order to guarantee the convergence of the operator product expansion (although this argument is subtle because we are varying the $(P_i, P_j, P_k)$ as complex variables independently).  This suggests that the function $C_0(P_i,P_j,P_k)$ can be {completely} determined by analyticity and simple physical constraints.  

We will now move on to the derivation of our result.  We begin in section \ref{sec:Cardy} with a detailed warm-up exercise, where we describe the derivation of various versions of Cardy's formula using the crossing kernel for modular transformations.  We then proceed to discuss the Moore-Seiberg procedure in more detail in section \ref{sec:MooreSeiberg}, before turning to the elementary crossing kernels in sections \ref{sec:fusion} and \ref{sec:modularS}.  We apply this to compute higher genus crossing kernels and OPE asymptotics in section \ref{sec:asymptoticFormulas}.  Large central charge limits, and comparisons to the literature, are discussed in section \ref{sec:semiclassical}.  Section \ref{sec:ETH} discusses the computation of the average value of the light-heavy-heavy OPE coefficients using the modular covariance of torus one-point funcitons. We relegate some details of the elementary crossing kernels and their asymptotics to the appendices.

\section{Cardy's formula from crossing kernels}\label{sec:Cardy}

To illustrate the main idea of the paper, we first revisit the derivation of the Cardy formula for primary states (and its large-spin version \cite{Collier:2018exn,Kusuki:2018wpa,Kusuki:2019gjs,Benjamin:2019stq,Maxfield:2019hdt}) using the modular S-matrix, a strategy which we will generalize in later sections. We follow the presentation and notation of \cite{Maxfield:2019hdt}, which contains some more details and applications. The relationship between the Cardy formula and the modular S-matrix was first elucidated in \cite{McGough:2013gka}.

\subsection{Natural variables for Virasoro representation theory}

As a preliminary, we introduce a parameterization of the CFT data that turns out to be natural for the representation theory of the Virasoro algebra. The central charge $c$ can be written in terms of a ``background charge'' $Q$ or ``Liouville coupling'' $b$ as
\begin{equation}
	c = 1+6Q^2 = 1+6(b+b^{-1})^2. 
\end{equation}
We will make the choice that $c>25$ corresponds to $0<b<1$, while $1<c<25$ corresponds to $b$ a pure phase in the first quadrant. To label Virasoro representations we use a variable $P$, or sometimes the equivalent $\alpha=\frac{Q}{2}-iP$, which is related to the more usually seen conformal weight by
\begin{equation}
	h = \left(\tfrac{Q}{2}\right)^2+P^2 =\alpha(Q-\alpha),
\end{equation}
and similarly $\bar{P}$ or $\bar{\alpha}$ in place of $\bar{h}$. Two things about this parameterisation should be noted. First, it is redundant, being invariant under the reflection reflections $P\rightarrow -P$ (or $\alpha\to Q-\alpha$). Secondly, it naturally splits unitary values of the weights ($h\geq 0$) into two distinct ranges: $h\ge\frac{c-1}{24}$ corresponds to real $P$ (or $\alpha\in \frac{Q}{2}+i\mathbb{R}$), and $0\leq h<\frac{c-1}{24}$, which corresponds to imaginary $P$ (or $\alpha\in(0,{Q\over 2})$).

\subsection{The partition function and density of primary states}

Now consider the torus partition function of a compact\footnote{By ``compact,'' we mean a CFT with a normalizable $SL(2,\mathbb{C})$-invariant vacuum state and a discrete spectrum of Virasoro primary operators.} CFT with $c>1$. The partition function encodes the spectrum of the theory, admitting a decomposition into Virasoro characters:
\begin{equation}
	Z(\tau,\bar\tau) = \chi_\id(\tau)\bar\chi_\id(\bar\tau) + \sum_i \chi_{P_i}(\tau)\bar{\chi}_{\bar{P}_i}(\bar\tau) 
\end{equation}
The sum runs over Virasoro primary states labelled by $i$, with conformal weights labelled by $P_i,\bar{P}_i$, and the nondegenerate Virasoro characters $\chi_P$ packaging together all states in a conformal multiplet are given by
\begin{equation}
	\chi_P(\tau) =\frac{q^{P^2}}{\eta(\tau)},
\end{equation}
where $q = e^{2\pi i\tau}$. The identity character $\chi_\id$ is distinguished because the corresponding representation is degenerate ($L_{-1}$ annihilates the vacuum state), so
\begin{equation}
	\chi_\id(\tau) = \chi_{\tfrac{i}{2}(b^{-1}+b)}(\tau)-\chi_{\tfrac{i}{2}(b^{-1}-b)}(\tau) = \frac{q^{-{Q^2\over 4}}(1-q)}{\eta(\tau)}.
\end{equation}
If there are any other conserved currents (operators with $h=0$ or $\bar{h}=0$) in the theory, we should similarly use this degenerate character for either the left- or right-moving half.

We can rewrite the character decomposition of the partition function in terms of a density of primary states $\rho$, writing
\begin{equation}\label{eq:pfrho}
	Z(\tau,\bar\tau) = \int \frac{dP}{2} \frac{d\bar{P}}{2} \, \rho(P,\bar{P}) \chi_P(\tau)\bar{\chi}_{\bar{P}}(\bar\tau),
\end{equation}
where $\rho$ is a distribution given by a sum of delta-functions $\delta(P-P_i)\delta(\bar{P}-\bar{P}_i)$ for each primary. Using the reflection symmetry, we make the choice that $\rho$ is an even distribution, so each primary contributes four terms related by reflections in $P,\bar{P}$, and we introduce the factors of $\frac{1}{2}$ in the integrals to avoid overcounting. It is also convenient to always use nondegenerate characters in the expansion, so for the identity (and other currents, if present), $\rho$ includes delta-functions with negative weight at $P,\bar{P}=\pm \frac{i}{2}\left(b^{-1}-b\right)$ to subtract the null descendants. Finally, we note that $\rho$ is a somewhat unconventional distribution, since it has support at imaginary values for operators with $h,\bar{h}<\frac{c-1}{24}$. This is nonetheless rigorously defined if we integrate against analytic test functions, of which the characters should form a complete set in an appropriate topology (see \cite{Maxfield:2019hdt} for more details).

\subsection{The modular S-transform}

Locality of a CFT implies invariance of the torus partition function under the modular S-transform, $Z(-1/\tau,-1/\bar\tau) = Z(\tau,\bar \tau)$, which in turn constrains the allowed CFT spectrum. We will reformulate this constraint directly on the density of states $\rho(P,\bar{P})$. To do this, first note that the modular S-transformation $\tau \to -1/\tau$ acts on individual characters as a Fourier (cosine) transform in the momentum:
\begin{equation}\label{eq:modularS}\begin{aligned}
\chi_{P}(-1/\tau) &= \int \frac{dP}{2}~ \chi_{P'}(\tau) \modS_{P'P}[\id] \\
\modS_{P'P}[\id] &= 2\sqrt{2}\cos(4\pi P P')
\end{aligned}\end{equation}
The kernel of this integral transform is the `modular S kernel' $\modS_{P'P}[\id]$, where the $[\id]$ label indicates that the partition function is a trivial example of the torus one-point function of the identity operator, with the generalization to nontrivial operators to follow. The notation emulates the situation in rational CFTs, where there are a finite number representations, so the modular kernel $\modS[\id]$ becomes a finite-dimensional matrix.

Given a function $Z(\tau,\bar{\tau})$ expanded in characters using a density of primary states as in \eqref{eq:pfrho}, we can take a modular S-transform and use the kernel \eqref{eq:modularS} to rewrite the transformed characters:
\begin{equation}
	Z(-1/\tau,-1/\bar{\tau}) = \int \frac{dP}{2}\frac{d\bar{P}}{2} \frac{dP'}{2}\frac{d\bar{P}'}{2} \modS_{P'P}[\id] \modS_{\bar{P}'\bar{P}}[\id] \rho(P,\bar{P})\chi_{P'}(\tau)\bar{\chi}_{\bar{P}'}(\bar{\tau})
\end{equation}
Exchanging order of integration between the primed and unprimed variables, we can interpret this as an expansion \eqref{eq:pfrho} of the modular transformed function with a transformed density of primary states:
\begin{equation}\label{eq:ModInvFT}
	\tilde{\rho}(P',\bar{P}') = \int \frac{dP}{2} \frac{d\bar{P}}{2}\; \modS_{P'P}[\id]\modS_{\bar{P}'\bar{P}}[\id] \rho(P,\bar{P})
\end{equation}
Since the partition function uniquely determines the spectrum, this equation expresses the modular S-transform as a Fourier transform acting on the density of primary states $\rho$.\footnote{We can strip off the characters since, by assumption, they are complete in the relevant space of test functions. This just means that a distribution is defined by its integral against all characters, i.e.~its corresponding partition function. The same applies for the more complicated transforms we encounter later.} In particular, a physical spectrum corresponding to a modular invariant theory is invariant under this Fourier transform:
\begin{equation}
	\text{Modular invariance} \iff  \tilde{\rho}(P,\bar{P})=\rho(P,\bar{P})
\end{equation}

From \eqref{eq:ModInvFT}, we can think of the modular S-matrix as the contribution of a single operator to the density of states in the transformed channel. The only exception to this is the degenerate representations with $h=0$ (or $\bar{h}=0$), so we introduce an `identity S-matrix'
\begin{equation}\label{eq:modularSVacuumLimit}
	\modS_{P\id}[\id] =\modS_{P,\frac{i}{2}(b^{-1}+b)}[\id]-\modS_{P,\frac{i}{2}(b^{-1}-b)}[\id] = 4\sqrt{2}\sinh(2\pi bP)\sinh(2\pi b^{-1}P),
\end{equation}
which encodes the contribution of such a degenerate state. The density of states $\modS_{P\id}[\id]\modS_{\bar{P}\id}[\id]$ dual to the vacuum will be of central importance for us.

\subsection{Cardy formulas}

The density of states $\rho(P,\bar{P})$ is a sum of delta-functions for each primary operator, so for a modular invariant spectrum, by taking the S-transform we can instead write it as a sum over modular S-matrices:
\begin{equation}\label{eq:sumOfS}
\rho(P,\bar{P}) = \modS_{P\id}[\id]\modS_{\bar{P}\id}[\id] + \sum_i  \modS_{PP_i}[\id]\modS_{\bar{P}\bar{P}_i}[\id]
\end{equation}
We have not explicitly included any nontrivial primary currents, which would contribute the identity S-matrix in $P$ and the nondegenerate S-matrix in $\bar{P}$ or vice versa. If such currents are present, it is most natural to organise the states into multiplets of an extended algebra, under which all currents are descendants of the vacuum, and use the modular S-matrix pertaining to the extended algebra.

Now consider this sum in the limit of large $P$ and/or $\bar{P}$. In this limit, the relative importance of the terms is determined by $P_i,\bar{P}_i$: for a state with $0<h<\frac{c-1}{24}$, the relevant S-matrix is exponentially suppressed relative to the vacuum:
\begin{equation}\label{eq:Sratio}
\frac{\modS_{PP'}[\id]}{\modS_{P\id}[\id]} \sim \begin{cases}
 	e^{-4\pi\alpha'P} & \alpha' = \tfrac{Q}{2}+iP' \in (0,\tfrac{Q}{2}) \\
 	2\cos(4\pi P P')e^{-2\pi Q P} &  P' \in \RR
 \end{cases}
 \quad\text{as }P\to\infty
\end{equation}
From this, we find (at least naively; we revisit this more carefully at the end of the section) that the density of states at large $P,\bar{P}$ asymptotically approaches the vacuum S-matrix:
\begin{equation}\label{eq:Cardy}
	\rho(P,\bar{P}) \sim \rho_0(P)\rho_0(\bar{P}) \text{ as }P,\bar{P}\to \infty, \text{ where } \rho_0(P) := \modS_{P\id}[\id] \sim \sqrt{2} e^{2\pi QP}
\end{equation}
This is of course nothing but Cardy's formula for the asymptotic density of primary states at large dimension, correct up to corrections exponential in $\sqrt{h},\sqrt{\bar{h}}$ coming from the lightest non-vacuum primary state.\footnote{This is the density in the $P,\bar{P}$ variables, so a Jacobian is required to convert to density in $h,\bar{h}$. For an asymptotic formula in dimension $\Delta=h+\bar{h}$ only, insensitive to spin, one simply integrates \eqref{eq:Cardy} over the circle $P^2+\bar{P}^2 = \Delta-\frac{c-1}{12}$, obtaining the Bessel function formulas of, for example, \cite{Mukhametzhanov:2019pzy,Collier:2017shs}.}

With this derivation, it becomes clear that the Cardy formula \eqref{eq:Cardy} is also valid in a `large spin' regime where we fix $h$ and take $\bar{h}\to\infty$ \cite{Collier:2018exn,Kusuki:2018wpa,Kusuki:2019gjs,Benjamin:2019stq,Maxfield:2019hdt}. In this limit, the relative suppression \eqref{eq:Sratio} of non-vacuum blocks is controlled by `barred' dimension only, so we require the additional assumption of a `twist gap' ($\bar{h}$ is bounded away from zero for all non-vacuum operators, so in particular there are no extra conserved currents). In this limit, for any fixed $h>\frac{c-1}{24}$, the density of states grows with spin $\ell$ as $e^{2\pi\sqrt{\frac{c-1}{6}\ell}}$, with a prefactor determined by $\rho_0(P)$; for any $h<\frac{c-1}{24}$, this prefactor is formally zero, which means that the density grows more slowly (perhaps still exponentially in $\sqrt{\ell}$, but with a smaller coefficient).

We therefore find that the asymptotic spectrum of CFTs is quite generally determined by the simple formula
\begin{equation}\label{eq:rho0}
	\rho_0(P) = \modS_{P\id}[\id] = 4\sqrt{2}\sinh(2\pi bP)\sinh(2\pi b^{-1}P),
\end{equation}
which we refer to as the `universal density of states' for $c>1$ compact CFTs without extended current algebras. Our derivation emphasizes that this object comes from the representation theory of the Virasoro algebra, describing the decomposition of the trivial representation after modular transformation.\footnote{In fact, $\rho_0$ has a purely representation theoretic characterization: it is the Plancherel measure on the space of representations of the Virasoro algebra \cite{Ponsot:1999uf}.\label{PlancherelFootnote}} In the remainder of the paper, we will show that another representation theoretic object similarly controls the OPE coefficients in a variety of limits.

Now, our argument for the asymptotic formula \eqref{eq:Cardy} was very imprecise, and indeed the result is simply false if interpreted literally, so we briefly discuss the sense in which it holds. The equation \eqref{eq:sumOfS} expressing the density of states as a sum of modular S kernels does not converge in the usual sense (and uniform convergence would be necessary for our argument to apply immediately), and since $\rho$ is a sum of delta functions, it does not have smooth asymptotic behaviour. Rather, the sum converges in the sense of distributions (it should converge when integrated against any test function), which requires some `smearing', and the the asymptotic formulas should be interpreted accordingly. The most conservative statement is that the formula applies in an integrated sense: the total number of states below a given energy or spin is asymptotic to the integral of the Cardy formula (see \cite{Mukhametzhanov:2019pzy,Ganguly:2019ksp,Pal:2019zzr} for a more detailed discussion and rigorous results). In the particular case of the Cardy formula, a very interesting recent paper \cite{Mukhametzhanov:2019pzy} has shown that if the averaging window is of fixed width in the large dimension limit, corrections due to the finite size of the averaging window only affect the order-one term in the expansion of the logarithm of the density of states at large dimension. For chaotic theories, we expect the far stronger statement that the asymptotic formula applies to a microcanonical density of states averaged over a small window (we require only that the window contains parametrically many states, so its width can shrink as fast as $e^{-S}$); this is a consequence of the eigenstate thermalization hypothesis (ETH) \cite{Srednicki:1994ne,deutsch1991quantum}. The exact interpretation of our asymptotic formulas is not the focus of this paper, so we will henceforth leave this aspect for future study.

\section{Crossing equations for general correlation functions}

We now extend this formulation of modular invariance as a transform on the density of states, discussed in section \ref{sec:Cardy}, to its most general context as a similar formulation of all consistency conditions of CFT$_2$.

\subsection{The Moore-Seiberg construction}\label{sec:MooreSeiberg}

In two dimensional CFTs, the most general correlation function of local operators, comprising $n$ operators $\op_1,\ldots,\op_n$ on a surface $\Sigma_g$ of arbitrary genus $g$ (which we denote by $G_{g,n}$), can be formulated entirely in terms of the basic data of the theory, namely the spectrum and OPE coefficients of primary operators.\footnote{This excludes correlation functions on surfaces with boundaries and/or nonorientable surfaces, both of which require additional data.} Note that this is far better than the situation in higher dimensions, where it is unclear how to determine general correlation functions, even on conformally flat manifolds such as the torus $(S^1)^d$, in terms of data of the theory on $\RR^d$. Here, we review the construction of general correlation functions, and the crossing relations required to consistently formulate the theory on an arbitrary surface.

The basic strategy is to break the surface into simple constituent pieces, separated by circular boundaries, and insert a complete set of states along each boundary. First, we insert a circle surrounding each operator insertion; by the state-operator correspondence, the operator insertion is equivalent to deleting a disc to produce a boundary, and projecting onto the corresponding state on that boundary. Label the resulting $n$ boundaries by an index $e\in\mathcal{E}$ (for `external') and let $k_e$ denote the operator on each boundary, falling in  Virasoro representations $P_{k_e},\bar{P}_{k_e}$.

 We are then left with a genus $g$ surface with $n$ boundaries, which we can decompose into $2g+n-2$ pairs of pants (that is, topological 3-holed spheres, occasionally called `trinions'), which we label by indices $t\in\mathcal{T}$, by cutting along a further $3g+n-3$ circles. Along each of these $3g+n-3$ `cuffs' where the pants are joined to one another, labelled by an index $i\in\mathcal{I}$ (for `internal'), we insert a complete set of states. Each term in the sum over states is then a product of amplitudes for each pair of pants, which can be conformally mapped to sphere three-point functions, and thus is fixed by the structure constants of the corresponding Virasoro primaries.

The contribution of descendants propagating along each cuff is completely fixed by Virasoro symmetry,  proportional to the OPE coefficients of the primaries from which they descend. We may therefore package together the contribution of all descendants of a particular set of primaries (labelled by $\{k_i\}_{i\in\mathcal{I}}$) together, into a `conformal block'. In other words, this is the sum over states described above, but restricting the states along each cuff $i$ to some chosen multiplet of the symmetry, in the representation $P_{k_i},\bar{P}_{k_i}$. By construction, the blocks are purely kinematic, depending on the surface $\Sigma_g$\footnote{The blocks (and the correlation functions) depend on the metric on the surface in two distinct ways. Firstly, there are finitely many moduli (the $3g+n-3$ complex parameters $\sigma$) determining the metric and operator locations up to equivalence under diffeomorphisms and Weyl transformations $g\mapsto e^{2\omega}g$, upon which the correlation function and blocks depend nontrivially. Secondly, there is the choice of metric within each such conformal class, which changes the correlation function only by kinematic factors: the conformal anomaly, and local conformal factors for each operator.} and the pair of pants decomposition\footnote{\label{DehnFoot}In fact, the decomposition into pairs of pants is not quite sufficient to determine the blocks. A Dehn twist, a relative rotation by angle $2\pi$ around a cuff, introduces phases $e^{2\pi i (h-c/24)}$ and $e^{-2\pi i (\bar{h}-\bar{c}/{24})}$ in $\mathcal{F}$ and $\bar{\mathcal{F}}$ respectively, so extra topological data is needed to keep track of these relative phases. When we combine blocks into the product $\mathcal{F}\bar{\mathcal{F}}$ with $c-\bar{c}\in 24\ZZ$ (here, we always have $c=\bar{c}$) and integer spin ($\bar{h}-h\in\ZZ$), this ambiguity cancels. We also require this extra data to fix an ambiguity in ordering of OPE coefficients, which pick up a sign under odd permutations of indices if the total spin is odd: $C_{\pi(1)\pi(2)\pi(3)}=\operatorname{sgn}(\pi)^{\ell_1+\ell_2+\ell_3} C_{123}$ for $\pi\in S_3$. Relatedly, note that the condition for unitarity is $C_{123}C_{321}\geq 0$, so for total odd spin $\ell_1+\ell_2+\ell_3$, $C_{123}$ is pure imaginary.}, the locations of operator insertions, the central charge, and the conformal weights $P_{k_e},\bar{P}_{k_e}$ and $P_{k_i},\bar{P}_{k_i}$ labelling the representations of the $n$ external and $3g+n-3$ internal operators. Since the conformal algebra factorizes into holomorphic and antiholomorphic sectors, the blocks also factorize in this way, so we can write them as a product $\mathcal{F}\bar{\mathcal{F}}$: $\mathcal{F}=\mathcal{F}[P_e](P_i|\sigma)$ depends on the $n$ external representations $P_e$ (for $e\in\mathcal{E}$), the $3g+n-3$ internal representations $P_i$ (for $i\in\mathcal{I}$), and kinematic variables collectively labelled by $\sigma$; we similarly have $\bar{\mathcal{F}}=\bar{\mathcal{F}}[\bar{P}_e](\bar{P}_i|\bar{\sigma})$. For Euclidean correlation functions, the kinematic variables $\sigma$ are (once a conformal frame has been specified) $3g-3+n$ complex numbers parameterising the complex structure moduli of $\Sigma_g$ and complex coordinates of the locations $x_e$ of operator insertions, and $\bar{\sigma}$ are complex conjugates of $\sigma$; more generally, $\sigma$ and $\bar{\sigma}$ need not be related in this way (for example, for Lorentzian kinematics they often become independent and real `lightcone' coordinates).

 The dynamical data of the theory appears through the spectrum of operators, and the OPE coefficients $C_{\partial t}$ for each pair of pants $t\in\mathcal{T}$, where $\partial t$ denotes a triple of indices $k_e$ or $k_i$ labelling the primary operators propagating in the three cuffs bounding $t$. The result is an expression of the following form for the correlation function:
\begin{equation}\label{eq:blockDecomp}
\begin{aligned}
	G_{g,n} &= \langle \op_1(x_1) \cdots \op_n(x_n) \rangle_{\Sigma_g} \\
	&= \underbrace{\sum_{k_{i=1}} \quad \cdots  \sum_{k_{i=3g+n-3}}}_{\text{Primaries on internal cuffs}}  \left(\prod_{t\in\mathcal{T}} C_{\partial t}\right) \mathcal{F}[P_{k_e}](P_{k_i}|\sigma)\bar{\mathcal{F}}[\bar{P}_{k_e}](\bar{P}_{k_i}|\bar{\sigma}) \\
	&= \int \left(\prod_{i\in\mathcal{I}} \frac{dP_{i}}{2}\right) \rho_\text{spec}[k_e](P_i,\bar{P}_i) \,\mathcal{F}[P_{k_e}](P_i|\sigma)\bar{\mathcal{F}}[\bar{P}_{k_e}](\bar{P}_i|\bar{\sigma}) \\
	 \rho_\text{spec}[k_e](P_i,\bar{P}_i) &= \sum_{k_i,i\in\mathcal{I}} \left(\prod_{t\in\mathcal{T}} C_{\partial t}\right) \prod_{i\in\mathcal{I}}\left(\delta(P_i-P_{k_i})\delta(\bar{P}_i-\bar{P}_{k_i}) +\text{(reflections)}\right)
\end{aligned}
\end{equation}
The last line defines a `spectral density' $\rho_\text{spec}$ analogous to the density of states in \eqref{eq:pfrho}, now with several internal operators, weighted by OPE coefficients; the `reflections' refers to an additional three terms with $P_{k_i}\to -P_{k_i}$ and/or $\bar{P}_{k_i}\to -\bar{P}_{k_i}$ so that $\rho_\text{spec}$ is an even function of these variables. This general case is rather abstract, but we will ultimately be interested in a few simple instances, for which we write concrete versions of \eqref{eq:blockDecomp} in later sections; for now, one illustrative example is shown in figure \ref{fig:G31ex}.
\begin{figure}
	\centering
	\includegraphics[width=.4\textwidth]{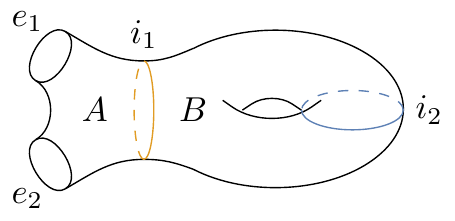}
	\caption[An example conformal block decomposition.]{A conformal block decomposition of the torus two-point function $G_{1,2}$, where the kinematic parameters $\sigma$ consist of a complex structure $\tau$ for the torus, and a separation $w$ between operators. We sum over representations in the internal cuffs; for the yellow cuff $i_1$, this corresponds to the operators appearing in the OPE of external operators $e_1,e_2$, and for the blue cuff $i_2$, an insertion of a complete set of states in the thermal trace. \\
	\begin{minipage}{\linewidth}
		\begin{equation*}
		G_{1,2}(\tau,w,\bar{\tau},\bar{w}) = \sum_{i_1}\sum_{i_2} C_{e_1e_2i_1}C_{i_1i_2i_2} \mathcal{F}[P_{e_1},P_{e_2}](P_{i_1},P_{i_2}|w,\tau) \bar{\mathcal{F}}[\bar{P}_{e_1},\bar{P}_{e_2}](\bar{P}_{i_1},\bar{P}_{i_2}|\bar{w},\bar{\tau})
	\end{equation*}
	\end{minipage}
	The OPE coefficients $C_{e_1e_2i_1}$, $C_{i_1i_2i_2}$ are associated with the pairs of pants labelled $A,B$ respectively, with $\partial A=(e_1,e_2,i_1)$ and $\partial B=(i_1,i_2,i_2)$.
	\label{fig:G31ex}}
\end{figure}

While our quick argument is sufficient to demonstrate that the conformal blocks exist, and are determined by Virasoro symmetry, it is another matter entirely to actually compute them. Closed form expressions are known only in very special cases. The most efficient way to compute them numerically is via recursion relations \cite{Zamolodchikov:1985ie,1987TMP....73.1088Z,Hadasz:2009db,Cho:2017oxl}, but even these are organised using different kinematic parameters and conformal frames for different channels, so it remains a challenging task to formulate crossing symmetry using them. The technical obstacles remain formidable even with the simplification of large central charge, where there are still few analytic results, and one must also confront the possibility of Stokes phenomena that are not well understood \cite{Faulkner:2017hll,Cardy:2017qhl,Collier:2018exn}. Fortunately, we will see later that for our purposes, it is not required to know anything about the blocks directly!

While we have a systematic procedure for constructing the correlation functions by sewing pairs of pants, it is far from unique, since there are infinitely many distinct ways to decompose a surface into pairs of pants. We refer to a choice of decomposition as a ``channel'', each channel giving rise to a corresponding conformal block decomposition of the correlation function. Consistency requires that the conformal block decompositions \eqref{eq:blockDecomp} give the same result for the correlation function, whichever channel we choose to use. This is a generalized statement of crossing symmetry or modular invariance, which imposes strong constraints on the data of the CFT.

To formulate this notion of crossing symmetry more directly in terms of the data of the CFT, we must first consider how to relate the block decompositions in different channels. Following the work of Moore and Seiberg \cite{Moore:1988qv,Moore:1988uz}, we can relate any two of the infinite collection of possible channels by repeated composition of a small number of elementary `moves', which can be described by purely topological relationships between pair of pants decompositions. We will make use of two such moves, `fusion' and `modular S' (or just S), illustrated and described in figure \ref{fig:ElementaryCrossingMoves}, along with an example where the two are composed.\footnote{For a complete set of moves, we also require `braiding', which acts on any two joined pairs of pants by adding a half twist to the separating cycle. The extra topological data required to fix the phases from footnote \ref{DehnFoot} is also necessary to uniquely prescribe the fusion/braiding moves among the infinitely many ways to split a sphere with four boundaries into two pairs of pants. It was only recently proved in \cite{Bakalov:rp} that fusion, braiding and S moves form a complete set of generators to relate any channels. We are grateful to Xi Yin for bringing \cite{Bakalov:rp} to our attention.}
\begin{figure}
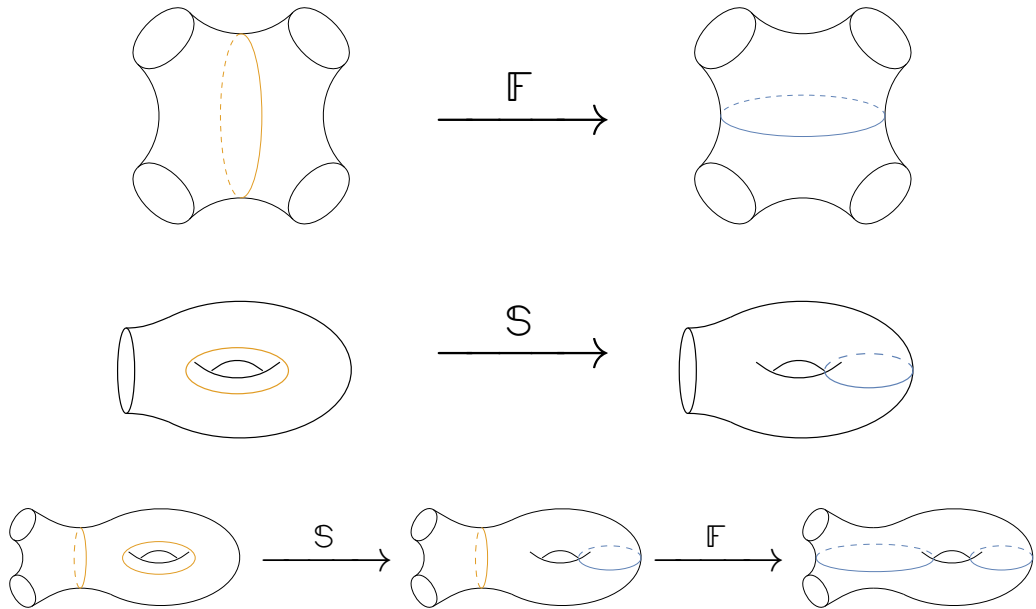

	\centering
	\includegraphics[width=.2\textwidth]{fourPointT} \raisebox{40pt}{\scalebox{2}{$\quad\xrightarrow{\quad\fusion\quad}\quad$}} \includegraphics[width=.2\textwidth]{fourPointS} \\
\vspace{20pt}
		\includegraphics[width=.2\textwidth]{torus1ptT} \raisebox{30pt}{\scalebox{2}{$\quad\xrightarrow{\quad\modS\quad}\quad$}} \includegraphics[width=.2\textwidth]{torus1ptS}\\
\vspace{20pt}
		\includegraphics[width=.2\textwidth]{torus2pt1} \raisebox{.034\textwidth}{\scalebox{1.5}{$\xrightarrow{\quad\modS\quad}$}} \includegraphics[width=.2\textwidth]{torus2pt2b}\raisebox{.034\textwidth}{\scalebox{1.5}{$\xrightarrow{\quad\fusion\quad}$}} \includegraphics[width=.2\textwidth]{torus2pt3}
	\caption{The elementary crossing moves relate different pair-of-pants decompositions of the four-punctured sphere and the once-punctured torus, or more generally anywhere that these appear as pieces of any decomposition of a surface. The associated crossing kernels relate Virasoro conformal blocks in the corresponding channels. The fusion kernel (top) relates sphere four-point Virasoro blocks in the S- and T-channels, and the modular kernel (middle) relates torus one-point blocks in modular $S$-transformed frames. In the final line, we show an example relating two channels in the torus two-point function $G_{1,2}$ by composing these moves.}\label{fig:ElementaryCrossingMoves}
\end{figure}

Now, we may informally think of the set of conformal blocks in any particular channel, labelled by the set of internal representations $\{P_i\}_{i\in \mathcal{I}}$, as forming a basis for correlation functions. Given a second channels, with a new set of internal cuffs $\mathcal{I}'$, there should be a change of basis matrix to the new variables $\{P_{i'}\}_{i'\in \mathcal{I}'}$, relating the two corresponding sets of blocks. From this point of view, it is plausible that the conformal blocks in any two channels can be related by an integral transform, with some `crossing kernel' $\kernel$:
\begin{equation}
	\mathcal{F}[P_e](P_{i}|\sigma) = \int \left(\prod_{i\in\mathcal{I}'}\frac{dP_{i'}}{2}\right) \mathcal{F}'[P_e](P_{i'}|\sigma') \kernel_{P_{i'} P_{i}}[P_e]
\end{equation}
We allow for a change of kinematic variables $\sigma\to\sigma'$ because natural variables (e.g.\ those appropriate for recursion relations) may be different in each channel. This equation is a generalisation of the relationship \eqref{eq:modularS} between characters in channels related by a modular transform, where the kernel $\kernel[P_e]$ was given by the modular S-matrix $\modS[\id]$. Furthermore, if we relate two channels by a composition of the elementary moves described above and in figure \ref{fig:ElementaryCrossingMoves}, the crossing kernel itself can be built by composing the kernels for the elementary moves.\footnote{For this, it is important that the same kernels apply for the elementary moves when the external operators are descendants of a given primary (which we sum over when these external legs become internal legs for a more complicated correlation function). This follows because descendant correlators can be obtained by acting with differential operators which are independent of the channel decomposition.} Remarkably, not only do these kernels exist, but for the elementary moves they are known in closed form! This is surprising and powerful when we consider how little analytic control we have regarding the conformal blocks. We will introduce these elementary kernels in the following subsections.

If the blocks are to be regarded as basis vectors, then the corresponding components of any particular correlation function are the OPE coefficients, as encoded in the spectral densities $\rho_\text{spec}$. Given a change of basis matrix $\kernel$, we can therefore relate the spectral densities in two channels by an integral transform with kernel $\kernel$, generalising \eqref{eq:ModInvFT}:\footnote{
This requires that the space of blocks is not overcomplete, so there is no nontrivial linear combination of blocks that gives the zero correlation function. This is extremely plausible; for example, the short distance behaviour of the correlator should be determined by the minimal dimension on which the spectral density has support.}
\begin{equation}
	\rho'_\text{spec}(P_{i'},\bar{P}_{i'}) = \int \left(\prod_{i\in\mathcal{I}} \frac{dP_{i}}{2}\frac{d\bar{P}_{i}}{2}\right) \kernel_{P_{i'}P_{i}}\kernel_{\bar{P}_{i'} \bar{P}_{i}} \rho_\text{spec}(P_{i},\bar{P}_{i})
\end{equation}
This is a direct statement of crossing or modular invariance, which makes no reference to the correlation function, the kinematics or the conformal blocks. As a corollary to the Moore-Seiberg construction, invariance under elementary moves implies invariance in complete generality, so four-point function crossing symmetry and torus one-point modular invariance for all operators suffice to prove consistency of a theory formulated on any surface. Nonetheless, more complicated correlation functions encode an infinite set of these constraints in a natural way, so more general crossing relations are still useful to learn about the theory, as we will see.

The elementary moves do not act freely on the space of channels, so they themselves are also highly constrained by the relations between moves. For example, we can consider a five-point function, made up of three pairs of pants, joined with two internal cuffs. Applying fusion moves alternately on each of the cuffs, we return to the original channel after five moves, and imposing that this combination of five $\fusion$'s acts trivially gives us the `pentagon identity' \eqref{eq:pentagon}, explained in more detail in section \ref{sec:LiouvilleEquations}. Assuming analyticity of the kernels, along with properties of degenerate representations, such identities suffice to determine the kernels uniquely \cite{Ponsot:1999uf,Ponsot:2000mt,Teschner:2001rv}.

The considerations we have described here have been understood and exploited for several decades, but largely in the context of rational models, for which only finitely many representations appear, so the kernel $\kernel$ is a finite-dimensional matrix (for a review, see \cite{AlvarezGaume:1989aq,AlvarezGaume:1989vk}). When applied to irrational theories, the technicalities are somewhat more subtle, and our aims must be more modest (we should certainly not hope to classify and solve all theories!), but this point of view nonetheless seems to be the most powerful way to formulate the constraints of crossing, even for irrational CFTs.

For the remainder of the section, we move beyond the abstract discussion to discuss more concretely the kernels for the elementary fusion and S moves, and their salient properties.

\subsection{Elementary crossing kernels 1: fusion}\label{sec:fusion}

The first of our elementary crossing moves arises when we consider the sphere four-point function
\begin{equation}
	G_{0,4}(z,\bar z) =\langle\mathcal{O}_1(0)\mathcal{O}_2(z,\bar z)\mathcal{O}_3(1)\mathcal{O}'_4(\infty)\rangle_{S^2},
\end{equation}
where $z,\bar z$ denote the conformal cross ratios. By successively taking the OPE between pairs of operators (corresponding to inserting a complete set of states in radial quantization), this can be written as sum over products of three-point functions of pairs of the external operators and intermediate operators:
\begin{equation}\label{eq:4ptrho}
\begin{aligned}
	G_{0,4}(z,\bar z) &= \sum_{\mathcal{O}_s}C_{12s}C_{34s}\,\mathcal{F}\sbmatrix{P_2 & P_1 \\ P_3 & P_4}(P_s|z)\bar{\mathcal{F}}\sbmatrix{\bar{P}_2 & \bar{P}_1 \\ \bar{P}_3 & \bar{P}_4}(\bar{P}_s|\bar z) \\
	&= \int \frac{dP_s}{2} \frac{d\bar{P}_s}{2}\, \rho_s(P_s,\bar{P}_s)\mathcal{F}\sbmatrix{P_2 & P_1 \\ P_3 & P_4}(P_s|z)\bar{\mathcal{F}}\sbmatrix{\bar{P}_2 & \bar{P}_1 \\ \bar{P}_3 & \bar{P}_4}(\bar{P}_s|\bar z),
\end{aligned}
\end{equation}
where $\mathcal{F}\sbmatrix{P_2 & P_1 \\ P_3 & P_4}(P|z)$ are the S-channel Virasoro blocks. In the second line we have written this decomposition as an integral against the S-channel `spectral density' $\rho_s$ (leaving implicit the dependence on external operators), which for a discrete spectrum is a sum of delta-functions weighted by the OPE coefficients $C_{12s}C_{34s}$; this is an example of the general decomposition \eqref{eq:blockDecomp}, analogous to \eqref{eq:pfrho} for the partition function.

For this expression, we have chosen to take the OPE between operators $\op_1$ and $\op_2$, giving the S-channel expansion (equivalently, we decompose the four-holed sphere into two pairs of pants, with cuffs $1,2,s$ and $s,3,4$). But the result must be the same if we instead choose to use the T-channel expansion, taking the OPE of operators $\op_2$ and $\op_3$. This associativity of the OPE leads to the crossing equation:
\begin{equation}
\begin{aligned}
	\int \frac{dP_s}{2} \frac{d\bar{P}_s}{2}\, \rho_s(P_s,\bar{P}_s)&\mathcal{F}\sbmatrix{P_2 & P_1 \\ P_3 & P_4}(P_s|z)\bar{\mathcal{F}}\sbmatrix{\bar{P}_2 & \bar{P}_1 \\ \bar{P}_3 & \bar{P}_4}(\bar{P}_s|\bar z)  \\&= \int \frac{dP_t}{2} \frac{d\bar{P}_t}{2}\, \rho_t(P_t,\bar{P}_t)\mathcal{F}\sbmatrix{P_2 & P_3 \\ P_1 & P_4}(P_t|1-z)\bar{\mathcal{F}}\sbmatrix{\bar{P}_2 & \bar{P}_3 \\ \bar{P}_1 & \bar{P}_4}(\bar{P}_t|1-\bar z)
\end{aligned}
\end{equation}
The T-channel spectral density $\rho_t$ appearing here is similar to $\rho_s$, but weighted by different OPE coefficients $C_{41t}C_{23t}$. This is the crossing relation between the two pair of pants decompositions of the four-holed sphere pictured on the top line of figure \ref{fig:ElementaryCrossingMoves}.

Continuing to follow the philosophy we applied to modular invariance in section \ref{sec:Cardy} and generalised in section \ref{sec:MooreSeiberg}, we will rewrite the crossing equation directly as a transform relating S- and T-channel spectral densities. To do this, we require an object expressing the decomposition of the T-channel Virasoro blocks in terms of S-channel blocks. This is the \emph{fusion kernel} (or crossing kernel, or $6j$ symbol), with the defining relation
\begin{equation}\label{eq:fusionTransformation}
	\mathcal{F}\sbmatrix{P_2 & P_3 \\ P_1 & P_4}(P_t|1-z) = \int \frac{d P_s}{2}\fusion_{P_sP_t}\sbmatrix{P_2 & P_1 \\ P_3 & P_4}\mathcal{F}\sbmatrix{P_2 & P_1 \\ P_3 & P_4}(P_s|z),
\end{equation}
which is analogous to the relation \eqref{eq:modularS} between the modular S-matrix and characters.

It is not a priori obvious that such an object should even exist, but it is a remarkable fact that it does, and an even more remarkable fact that it has been explicitly constructed by Ponsot and Teschner \cite{Ponsot:1999uf,Ponsot:2000mt,Teschner:2001rv}. A closed form expression is given in (\ref{eq:explicitFusion}) in appendix \ref{app:explicitForms}, which contains the necessary technical results, many of which were derived in \cite{Collier:2018exn}. We discuss the most relevant properties in a moment.

With the fusion kernel $\fusion$ in hand, we can now write the crossing equation as a transform relating the spectral density in each channel, just as in \eqref{eq:ModInvFT}:
\begin{equation}\label{eq:fusionTransform}
	 \rho_s(P_s,\bar{P}_s) = \int \frac{d P_t}{2}\frac{d \bar{P}_t}{2} \,\fusion_{P_sP_t} \fusion_{\bar{P}_s \bar{P}_t} \rho_t(P_t,\bar{P}_t)
\end{equation}
Here we have suppressed the notation labelling the external operators, but it should be borne in mind that the kernel of this transform depends on the external operator dimensions $P_{1,2,3,4}$.\footnote{There is a similar transform to write the S-channel spectral density in terms of U-channel data (with density weighted by OPE coefficients $C_{13u}C_{u24}$) using the braiding kernel. This is a fusion kernel conjugated by phases, which become signs for integer spins:
	\begin{equation}
		\rho_s(P_s,\bar{P}_s) = \int \frac{d P_u}{2}\frac{d \bar{P}_u}{2} \,(-1)^{\ell_1+\ell_4+\ell_u+\ell_s}\fusion_{P_sP_u}\sbmatrix{P_2 & P_	1 \\ P_4 & P_3} \fusion_{\bar{P}_s \bar{P}_u}\sbmatrix{P_2 & P_1 \\ P_4 & P_3} \rho_u(P_u,\bar{P}_u)
	\end{equation}
	The resulting signs for odd spins are much the same as for U-channel inversion in \cite{Caron-Huot:2017vep}, for example.}

Like the modular transform of the vacuum \eqref{eq:rho0} was the most important object in section \ref{sec:Cardy}, the fusion transform of the vacuum will play a correspondingly central role for our new asymptotic formulas. This can only appear in the case that the external operator dimensions are equal in pairs, $P_1=P_4$ and $P_2=P_3$ (in the T-channel). In that case, the fusion kernel simplifies\footnote{Unlike for the modular S-matrix in section \ref{sec:Cardy}, the fusion kernel for the identity can be obtained as a continuous $h_t\to0$ limit of the generic fusion kernel (with external operators identical in pairs). This occurs because the null states continuously decouple (their OPE coefficients go to zero continuously as $h_t\to 0$). See footnote \ref{foot:modS} for a more detailed comparison.} \cite{Collier:2018exn}, and we find it convenient to write it as
\begin{equation}\label{eq:idFusion}
	\fusion_{P_s\id}\sbmatrix{P_2 & P_1 \\ P_2 & P_1} = \rho_0(P_s) C_0(P_1,P_2,P_s),
\end{equation}
where $\rho_0(P)$ is the density of states appearing as the modular S-transform of the vacuum \eqref{eq:rho0}. It turns out that $C_0$ is then symmetric under the exchange of all three of its arguments, and has a simple explicit expression in terms of the special function $\Gamma_b$:
\begin{equation}\label{eq:C0}
	C_0(P_1,P_2,P_3) = \frac{1}{\sqrt{2}}{\Gamma_b(2Q)\over \Gamma_b(Q)^3}\frac{\prod_{\pm\pm\pm}\Gamma_b\left(\tfrac{Q}{2}\pm iP_1\pm iP_2 \pm iP_3\right)}{\prod_{k=1}^3\Gamma_b(Q+2iP_k)\Gamma_b(Q-2iP_k)}
\end{equation}
The $\prod$ in the numerator denotes the product of the eight combinations related by the reflections $P_k\to -P_k$. The function $\Gamma_b$ is a `double' gamma function, which is meromorphic, with no zeros, and with poles at argument $-mb-nb^{-1}$ for nonnegative integers $m,n$ (similarly to the usual gamma function, which has poles at nonpositive integers).

If external operators are sufficiently light (specifically, $\alpha_1+\alpha_2\leq \frac{Q}{2}$ or $\alpha_3+\alpha_4\leq \frac{Q}{2}$), the fusion kernel has a new subtlety, arising from poles in $P_s$ that cross the real axis. In order to maintain analyticity in the parameters, the contour in the decomposition (\ref{eq:fusionTransformation}), which is implicitly taken to run along the real $P_s$ axis, must be deformed. We can take the deformed contour to run along the real $P_s$ axis, but must additionally include circles surrounding the poles which have crossed the axis, contributing residues. This gives rise to a finite sum of S-channel operators with imaginary $P_s$ ($h_s<\frac{c-1}{24}$) in the decomposition of the T-channel conformal block. See \cite{Collier:2018exn} for more details. We can describe this by including a sum of $\delta$-functions supported at imaginary $P_s$ in the kernel $\fusion$ \cite{Maxfield:2019hdt}.

The non-vacuum kernels with T-channel dimension $h_t>0$ will be important for us only to compare their asymptotic contribution to the S-channel. The key result, established in \cite{Collier:2018exn}, is precisely analogous to \eqref{eq:Sratio} for the modular S-matrix:
\begin{equation}\label{eq:fusionSuppression}
\frac{\fusion_{P_s P_t}}{\fusion_{P_s\id}} \approx \begin{cases}
 	e^{-2\pi\alpha_t P_s} & \alpha_t = \tfrac{Q}{2}+iP_t \in (0,\tfrac{Q}{2}) \\
 	e^{-\pi Q P_s}\cos(2\pi P_t P_s) &  P_t \in \RR
 \end{cases}
 \quad\text{as }P_s\to\infty
\end{equation}
This result is accurate up to a factor independent of $P_s$, see equation (\ref{eq:fusionSuppression}). 

\subsection{Elementary crossing kernels 2: modular S}\label{sec:modularS}

The second elementary move is a modular transform applied to one-point functions of Virasoro primary operators on the torus
\begin{equation}
G_{1,1}(-\tau,\bar\tau) = \langle \mathcal{O}_0\rangle_{T^2(\tau,\bar \tau)},
\end{equation}
where $\tau$ labels the complex structure of the torus, and the conformal weight of the external operator is $h_0  = \left({Q\over 2}\right)^2+P_0^2= \alpha_0(Q-\alpha_0)$. The translation invariance of the torus means that the correlation function is independent of the location of the operator.

Generalizing the modular invariance of the torus partition function (which is the special case where the external operator $\op_0$ is the identity), $G_{1,1}$ transforms covariantly under modular transformations, in particular the S-transform $\tau \rightarrow -1/\tau$:
\begin{equation}
	G_{1,1}(-1/\tau,-1/\bar\tau) = \tau^{h_0}\bar{\tau}^{\bar{h}_0}G_{1,1}(\tau,\bar\tau)
\end{equation}
The factor $\tau^{h_0}\bar{\tau}^{\bar{h}_0} = |\tau|^\Delta e^{-i\ell_0 \arg\tau}$ comes from rescaling and rotating the torus so the thermal circle becomes the spatial circle\footnote{Performing this transform twice corresponds to rotating the torus through an angle $\pi$ and gives a factor $(-1)^{\ell_0}$, from which we conclude that $G_{1,1}$ is zero for operators with odd spin, since any nonzero expectation value would break this $\ZZ_2$ symmetry.}. It occurs because the definition of the one-point function implicitly makes a choice of metric on the torus, namely the flat metric in which the spatial circle has length $2\pi$; after modular transform, the cycle interpreted as the spatial circle changes, and hence the metric is rescaled. The discussion of subsection \ref{sec:MooreSeiberg} implicitly assumed that we use the same metric for every channel, so there were no such factors.

We can write this correlation function in terms of the usual CFT data by inserting a complete set of states on the spatial circle, and collecting the contributions from each Virasoro representation into torus conformal blocks $\mathcal{F}[P_0](P|\tau)$ with internal primary weight $P$:\footnote{Explicitly, $\mathcal{F}[P_0](P|\tau)=\Tr_{\!P}(e^{2\pi i \tau L_0}\op_0)$, where the trace is taken over the representation of the Virasoro algebra with weight labelled by $P$, normalising the expectation value of $\op_0$ in the lowest weight state to unity.}
\begin{equation}\begin{aligned}
	G_{1,1}(\tau,\bar\tau) &= \sum_{\mathcal{O}}C_{\mathcal{O}\mathcal{O}\mathcal{O}_0}\mathcal{F}[P_0](P|\tau)\bar{\mathcal{F}}[\bar{P}_0](\bar{P}|\bar\tau) \\
	&= \int \frac{dP}{2} \frac{d\bar{P}}{2} \rho[\op_0](P,\bar{P}) \mathcal{F}[P_0](P|\tau)\bar{\mathcal{F}}[\bar{P}_0](\bar{P}|\bar\tau)
\end{aligned}\end{equation}
In the second line we have defined the thermal spectral density $\rho[\op_0]$ for the external operator $\op_0$, consisting of $\delta$-functions for each internal operator with coefficient $C_{\mathcal{O}\mathcal{O}\mathcal{O}_0}$, analogously to \eqref{eq:pfrho} and \eqref{eq:4ptrho}, and another special case of \eqref{eq:blockDecomp}.

Reprising the same strategy, we will recast modular covariance as invariance of $\rho_{\op_0}(P,\bar{P})$ under an S-transform, directly generalizing \eqref{eq:ModInvFT} for the density of states. To do this, we introduce the torus one-point kernel, the object which decomposes torus one-point conformal blocks into the modular-$S$ transformed frame:
\begin{equation}\begin{aligned}\label{eq:modularSTransform}
	\tau^{h_0}\mathcal{F}[P_0](P|\tau) &= \int \frac{d P'}{2}\mathcal{F}[P_0](P'|-1/\tau) \modS_{P'P}[P_0]
\end{aligned}\end{equation}
Given this object, the modular S transformation acts on the spectral density as
\begin{equation}\label{eq:modularScov}
	\tilde{\rho}[\op_0](P',\bar{P}') = \int \frac{dP}{2} \frac{d\bar{P}}{2} \modS_{P'P}[P_0]\modS_{\bar{P}'\bar{P}}[\bar{P}_0] \rho[\op_0](P,\bar{P}),
\end{equation}
and modular covariance of $G_{1,1}$ is stated as $\tilde{\rho}[\op_0]=\rho[\op_0]$.

Once again, we are fortunate to have an explicit expression for the modular S-kernel due to Teschner \cite{Teschner:2003at} (see also \cite{Nemkov:2015zha,Nemkov:2016ikx}). We reproduce the precise formula in (\ref{eq:explicitModularS}) of appendix \ref{app:explicitForms}, where we demonstrate various important properties of the kernel, the most salient of which we now state.

Most important for us is that, like the fusion kernel, the modular S-kernel simplifies when the external operator is the identity, taking $h_0\to 0$ ($P_0\to i\frac{Q}{2}$). In this limit, we find that

\begin{equation}\begin{aligned}\label{eq:ModularSMatrix}
	\modS_{PP'}[P_0]\to \modS_{PP'}[\id] &= 2\sqrt{2}\cos\left(4\pi P P'\right),
\end{aligned}\end{equation}
recovering the modular S-matrix for non-degenerate torus characters (\ref{eq:modularS}) from section \ref{sec:Cardy}. Note that the kernel relevant for inversion of the vacuum character, namely
\begin{equation}
\label{eq:vacuumModularSMatrix}
	\modS_{P\id}[\id] 
	= 4\sqrt{2}\sinh(2\pi bP)\sinh(2\pi b^{-1}P)
\end{equation}
as in equation (\ref{eq:modularSVacuumLimit}), is not recovered by a straightforward $\alpha'\to 0$ limit of (\ref{eq:ModularSMatrix}), because the degenerate vacuum character is not given simply by the $h'\to 0$ limit of the non-degenerate character. This is unlike the fusion kernel, where the identity kernel is obtained by an $\alpha_t\to 0$ limit of the generic kernel with external operators identical in pairs: in that case the null descendants continuously decouple in the $h_t\to 0$ limit.\footnote{Since $\frac{\langle h'|L_1 \op_0 L_{-1}|h'\rangle}{\langle h'|L_1 L_{-1}|h'\rangle} = \frac{2h'+h_0(h_0-1)}{2h'}\langle h'|\op_0|h'\rangle$, we can take a vacuum limit in which the null descendant is decoupled by fixing $h'=-\frac{1}{2}h_0(h_0-1)\sim \frac{1}{2}h_0$ and taking $h_0\to 0$. Indeed, taking a limit $\alpha_0,\alpha'\to 0$ with $\alpha'\sim \frac{1}{2}\alpha_0$, one can explicitly check that $\modS_{PP'}[P_0]\to \modS_{P\id}[\id]$ (for a derivation, see \eqref{eq:SvacLimit} and surrounding discussion). In contrast, for the fusion kernel we can take a more direct limit because the matrix elements $\frac{\langle h_1|\op_1 L_{-1}|h_t\rangle\langle h_t|L_1\op_2 |h_2\rangle}{\langle h_t|L_1 L_{-1}|h_t\rangle} = \frac{h_t}{2}\langle h_1|\op_1 |h_t\rangle\langle h_t|\op_2 |h_2\rangle$ go to zero as $h_t\to 0$.\label{foot:modS}}

The second important property for us will be the behaviour of the kernel in the large dimension limit $P\to \infty$, which we normalise by the vacuum S-matrix $\modS_{P\id}[\id]\approx e^{2\pi QP}$ for comparison:
\begin{equation}\label{eq:modularSAsymptotics}
\frac{\modS_{PP'}[P_0]}{\modS_{P\id}[\id]} \approx \begin{cases}
 	e^{-4\pi\alpha'P}P^{h_0} & \alpha' = \tfrac{Q}{2}+iP' \in (0,\tfrac{Q}{2}) \\
 	e^{-2\pi Q P}\cos(4\pi P P') P^{h_0} &  P' \in \RR
 \end{cases}
 \quad\text{as }P\to\infty
\end{equation}
These formulas, derived in appendix \ref{app:torusOnePtAsymptotics}, are accurate up to a constant (that is, independent of $P$) factor. Crucially, this ratio is exponentially suppressed at large $P$, as long as $h'>0$. This result reduces to \eqref{eq:Sratio} when the external operator is the identity.

\section{OPE asymptotics from crossing kernels}\label{sec:asymptoticFormulas}

Now that we have formulated the consistency conditions as statements about transforms of spectral densities, it is simple to repeat the arguments of section \ref{sec:Cardy}, which led to the Cardy formula, in a variety of new situations. Specifically, we study crossing for the three correlation functions which decompose into two pairs of pants, and extract asymptotic formulas for squares of OPE coefficients.

\subsection{Sphere four-point function: heavy-light-light}

For our first example, we study the constraints of crossing symmetry for the four-point function of pairwise identical operators. We have already introduced all the required definitions and results in subsection \ref{sec:fusion}; in particular, we have the fusion transformation \eqref{eq:fusionTransformation} relating S- and T-channel spectral densities,
\begin{equation}
	 \rho_s(P_s,\bar{P}_s) = \int \frac{d P_t}{2}\frac{d \bar{P}_t}{2} \,\fusion_{P_sP_t} \fusion_{\bar{P}_s \bar{P}_t} \rho_t(P_t,\bar{P}_t),
\end{equation}
and the result \eqref{eq:fusionSuppression} that the fusion kernel for operators of positive dimension $h_t>0$ is exponentially suppressed compared to the identity at large $P_s$. This is precisely the same situation we had for the modular S-matrix when we derived the Cardy formula \eqref{eq:Cardy}, so repeating that argument gives us an analogous result for the S-channel spectral density:
\begin{equation}
	\rho_s(P_s,\bar{P}_s) \sim \fusion_{P_s\id}\sbmatrix{P_2 & P_1 \\ P_2 & P_1} \fusion_{\bar{P}_s\id}\sbmatrix{ \bar{P}_2 & \bar{P}_1 \\ \bar{P}_2 & \bar{P}_1 },\quad P_s,\bar{P}_s \to \infty.
\end{equation}
This finding is not new, but was one of the main results of \cite{Collier:2018exn}. The focus of that paper was the large spin limit of fixed $P_s$ and $\bar{P}_s\to\infty$, but we here emphasise that this also holds for large dimension (both $P_s,\bar{P}_s\to\infty$), in fact more generally since we need not assume existence of a twist gap in that case.

In higher dimensional CFTs, the analogous operation of expanding the T-channel identity block (which is simply the product of two-point functions) into the S-channel defines the spectrum and OPE coefficients of `double trace' operators of mean field theory (MFT). The identity fusion kernel can therefore be thought of as a deformation of MFT to include Virasoro symmetry, and the corresponding spectral data was accordingly dubbed ``Virasoro mean field theory'' (VMFT) in \cite{Collier:2018exn}. The large-spin universality of the identity kernel is the $d=2$ analogue of the result for $d>2$ that there exist `double-twist' operators whose dimensions and OPE coefficients approach those of MFT at large spin \cite{Komargodski:2012ek,Fitzpatrick:2012yx}.

The analogy with double-twist operators in higher dimensions is sharpest for $h<\frac{c-1}{24}$. If the external operators $\op_1,\op_2$ have sufficiently low twist, then there are a finite number of trajectories that asymptote at large spin to discrete values of $h<\frac{c-1}{24}$; see \cite{Collier:2018exn} for details. There is also a continuum starting at $h = \frac{c-1}{24}$ described by the smooth VMFT OPE density, which has no known analog in higher dimensions.

For $h>\frac{c-1}{24}$, either fixed in the large spin limit or taken to be large simultaneously with $\bar{h}$, the asymptotic spectrum encoded in the fusion kernel is a smooth function of $P,\bar{P}$. Just as for the Cardy formula explained in section \ref{sec:Cardy}, \eqref{eq:4ptrho} should then be interpreted as a microcanonical statement about the asymptotic spectral density integrated over a window of energies. We can translate the result to a microcanonical average of OPE coefficients, by dividing by the Cardy formula \eqref{eq:Cardy} giving the asymptotic density of primary states $\rho(P_s,\bar{P_s})\sim\rho_0(P_s)\rho_0(\bar{P}_s)$ in the relevant limits. Writing the identity fusion kernel in the form \eqref{eq:idFusion} of the universal density $\rho_0(P_s)$ times $C_0(P_1,P_2,P_s)$, we find that $C_0$ gives the microcanonical average of the OPE coefficients:
\begin{equation}
	\overline{|C_{12s}|^2} \sim C_0(P_1,P_2,P_s)C_0(\bar{P}_1,\bar{P}_2,\bar{P}_s),~P_s,\bar{P}_s \to \infty.
\end{equation}
This result is valid for any two fixed operators $\op_1,\op_2$, averaging over operators $\op_s$ in either a large dimension or large spin limit.

The asymptotic form of $C_0$ in this limit was computed in \cite{Collier:2018exn}:
\begin{equation}\label{eq:C0HLL}
	C_0(P_1,P_2,P_s) \sim 2^{-4P_s^2}e^{-\pi Q P_s} P_s^{4(h_1+h_2)-{3Q^2+1\over 2}}{2^{Q^2-2\over 6}\Gamma_0(b)^6\Gamma_b(2Q)\over\Gamma_b(Q)^3\Gamma_b(Q+2iP_1)\Gamma_b(Q-2iP_1)\Gamma_b(Q+2iP_2)\Gamma_b(Q-2iP_2)},
\end{equation}
where $\Gamma_0(b)$ is a special function that appears in the large-argument asymptotics of $\Gamma_b$; see appendix A of \cite{Collier:2018exn} for more details. 
The first factor exactly cancels a similar factor in the conformal blocks ($\mathcal{F}\approx (16q)^{h_s}$ \cite{Zamolodchikov:1985ie}), ensuring that the block expansion has the correct domain of convergence. A formula of this form for the asymptotics of the averaged heavy-light-light structure constants was first obtained in \cite{Das:2017cnv}.
In that paper, the authors used the asymptotics of the Virasoro four-point blocks in the heavy limit $h_s\to\infty$ \cite{Zamolodchikov:1985ie}, subsequently taking a $z\to 1$ limit to reproduce the OPE singularity from the T-channel identity operator. Their result matches the leading asymptotics of our formula \eqref{eq:C0HLL} when written in terms of the conformal weights and central charge (as in equation \eqref{eq:HLL}); we find new terms appearing at subleading order arising from a subtlety in the order of $h_s\to\infty$ and $z\to 1$ limits. Working directly with the spectral densities allows us to avoid such difficulties in studying conformal blocks.

\subsection{Torus two-point function: heavy-heavy-light}
For our second example, we study the two-point function of identical Virasoro primaries on the torus:
\begin{equation}
	G_{1,2}(\tau,\bar \tau;w,\bar w) = \langle \mathcal{O}_0(w,\bar w)\mathcal{O}_0(0,0)\rangle_{T^2(\tau,\bar\tau)}
\end{equation}
There are two qualitatively distinct ways to decompose such a correlation function into conformal blocks. Firstly, we can take the OPE between the two operators and insert a single complete set of states around a cycle of the torus, which we call the OPE channel. Secondly, we can insert two complete sets of states between the operators on each side of the thermal circle, which we call the necklace channel.
\begin{equation}\begin{aligned}
	G_{1,2}(\tau,\bar \tau;w,\bar w) &= \sum_{\mathcal{O}_1}\sum_{\mathcal{O}_2}|C_{012}|^2\mathcal{F}^{(\text{N})}[P_0](P_1,P_2|q_1,q_2)\bar{\mathcal{F}}^{\text{(N)}}[\bar{P}_0](\bar{P}_1,\bar{P}_2|\bar{q}_1,\bar{q}_2)\\
	&= \int \frac{dP_1}{2}\frac{d\bar{P}_1}{2}\frac{dP_2}{2}\frac{d\bar{P}_2}{2} \rho_\text{N}(P_1,P_2,\bar{P}_1,\bar{P}_2)\mathcal{F}^{(\text{N})}[P_0](P_1,P_2|q_1,q_2)\bar{\mathcal{F}}^{\text{(N)}}[\bar{P}_0](\bar{P}_1,\bar{P}_2|\bar{q}_1,\bar{q}_2)\\
	&= \sum_{\mathcal{O}_1'}\sum_{\mathcal{O}'_2}C_{002'}C_{2'1'1'}\mathcal{F}^{(\text{OPE})}[P_0]( P_1', P_2'|q,v)\bar{\mathcal{F}}^{(\text{OPE})}[\bar{P}_0](\bar{P}_1',\bar{P}_2'|\bar q,\bar v)\\
	&= \int \frac{dP_1'}{2}\frac{d\bar{P}_1'}{2}\frac{dP_2'}{2}\frac{d\bar{P}_2'}{2} \rho_\text{OPE}(P_1',P_2',\bar{P}_1',\bar{P}_2')\mathcal{F}^{(\text{OPE})}[P_0]( P_1', P_2'|q,v)\bar{\mathcal{F}}^{(\text{OPE})}[\bar{P}_0](\bar{P}_1',\bar{P}_2'|\bar q,\bar v)
\end{aligned}\end{equation}
The second and fourth lines define `necklace' and `OPE' spectral densities $\rho_\text{N}$, $\rho_\text{OPE}$. We have written the blocks using different kinematic variables, since the natural parameters (for recursion relations, for example \cite{Cho:2017oxl}) are different in the two channels. In the necklace channel, $q_1$ and $q_2$ encode a Euclidean time evolution, between the two operator insertions, and then round the torus back to the first operator insertion; in the OPE channel, there is only one such parameter $q$, along with a separation $v$ between the operators controlling the OPE. These parameters can be related to one another, but all our results are derived without explicit reference to any kinematics.

\begin{figure}
		\qquad\qquad\, \includegraphics[width=.2\textwidth]{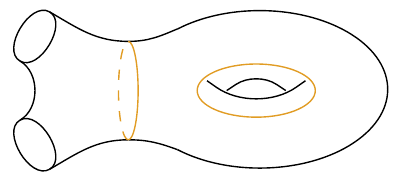}  \raisebox{.034\textwidth}{\scalebox{1.5}{$=\int \frac{dP_1}{2}\modS_{P_1P_1'}[P'_2]$}} \includegraphics[width=.2\textwidth]{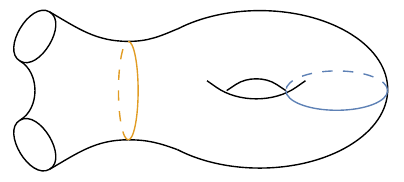} \\
		\flushright\raisebox{.034\textwidth}{\scalebox{1.5}{$=\int \frac{dP_1}{2}\frac{dP_2}{2}\modS_{P_1P_1'}[P'_2] \fusion_{P_2P_2'}\sbmatrix{P_0&P_0\\P_1&P_1}$}}
		\includegraphics[width=.2\textwidth]{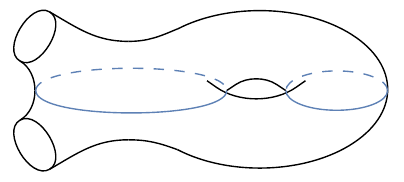}
	\caption{The sequence of Moore-Seiberg moves to express the OPE channel torus two-point block in terms of necklace channel blocks: a modular S, followed by a fusion move. \label{fig:torus2pt}}
\end{figure}

We will consider the crossing kernel that decomposes torus two-point blocks for identical operators in the OPE channel (with internal Liouville momenta $P'_1,P'_2$) into two-point blocks in the necklace channel (in the modular $S$-transformed frame). This sewing procedure is illustrated in figure \ref{fig:torus2pt}, from which we see that the required kernel is simply given by the product of the torus one-point kernel and the sphere four-point kernel:
\begin{equation}\begin{gathered}
	\kernel_{P_1P_2;P'_1P'_2}[P_0] = \modS_{P_1P'_1}[P'_2]\fusion_{P_2 P_2'} \sbmatrix{P_0 & P_1 \\ P_0 & P_1} \\
	\rho_\text{N}(P_1,P_2,\bar{P}_1,\bar{P}_2) = \int\frac{dP_1'}{2}\frac{d\bar{P}_1'}{2}\frac{dP_2'}{2}\frac{d
	\bar{P}_2'}{2} \kernel_{P_1P_2;P'_1P'_2}[P_0]\kernel_{\bar{P}_1\bar{P}_2;\bar{P}'_1\bar{P}'_2}[\bar{P}_0]\rho_\text{OPE}(P_1',P_2',\bar{P}_1',\bar{P}_2')
\end{gathered}\end{equation}

In an appropriate limit, the necklace channel data will be dominated by the identity propagating in both internal cuffs of the OPE channel, described by the identity kernel
\begin{equation}\begin{aligned}
	\kernel_{P_1P_2;\id\id}[P_0] &= \modS_{P_1\id}[\id]\fusion_{P_2\id}\sbmatrix{P_0& P_1 \\ P_0 & P_1}\\
&= \rho_0(P_1)\rho_0(P_2)C_0(P_0,P_1,P_2).
\end{aligned}\end{equation}
Once again, the asymptotics of $C_0$ universally governs the asymptotics of OPE coefficients, this time in a `heavy-heavy-light' limit, where one operator is fixed, and the other two operators are taken to have large dimensions. Corrections to this identity contribution due to the exchange of non-vacuum primaries in the OPE channel are exponentially suppressed when we take $P_1,P_2$ to be large, just as we have seen before. The technical result required to show this is
\begin{equation}\label{eq:torusTwoPointId}
	\frac{\kernel_{P_1P_2;P'_1P'_2}[P_0]}{\kernel_{P_1P_2;\id\id}[P_0]} \approx e^{-2\pi\alpha_1'P_1}
\end{equation}
in the limit $P_1,P_2\to\infty$, with either the ratio or difference of $P_1$ and $P_2$ held fixed. This result is asymmetric in $P_1$ and $P_2$ because the OPE channel does not treat operators symmetrically\footnote{We could make the derivation symmetric in $P_1,P_2$ by including an extra fusion move, so that we are relating two different OPE channels. Starting with the identity block, this extra fusion move is `free' (that is, the necklace identity block is equal to the OPE identity block), since there are external operators for $\fusion$ in the identity representation. However, this extra move makes the argument for suppression of non-vacuum operators more technically challenging.}; it guarantees suppression of all non-vacuum blocks because $\alpha_2'$ cannot be nonzero unless $\alpha_1'$ is also nonzero. See the discussion in appendix \ref{app:higherGenusAsymptoticCorrections} for more details. 

As in the case of the sphere four-point function, this result means that the necklace channel spectral density is well approximated by exchange of the vacuum Verma module in the OPE channel when the internal weights are taken to be heavy:
\begin{equation}
	\rho_{\text{necklace}}^{(P_0,\bar{P}_0)}(P_1,\bar{P}_1;P_2,\bar{P}_2) \approx \kernel_{P_1P_2;\id\id}[P_0]\kernel_{\bar{P}_1\bar{P}_2;\id\id}[\bar{P}_0],~P_1,P_2,\bar{P}_1,\bar{P}_2\to \infty
\end{equation}
Thus the kernel corresponding to propagation of the identity in the OPE channel (\ref{eq:torusTwoPointId}) encodes an asymptotic formula for OPE coefficients in the heavy-heavy-light regime, averaged over the heavy operators, and for any fixed light operator. Stripping off the density of states of the heavy operators, we have
\begin{equation}
	\overline{\left|C_{012}\right|^2} \sim C_0(P_0,P_1,P_2) C_0(\bar{P}_0,\bar{P}_1,\bar{P}_2),~P_1,P_2,\bar{P}_1,\bar{P}_2\to \infty.
\end{equation}
As in the case of the sphere four-point function, in the presence of a nonzero twist gap the above asymptotic formula also holds in the large-spin regime when only $P_1,P_2$ or $\bar{P}_1,\bar{P}_2$ are taken to be large.

Now that there are multiple internal weights, there are several distinct ways to take the large-weight limit. First, we can take the weights to infinity at fixed ratio $\frac{P_2}{P_1}$, assuming without loss of generality that $P_1>P_2$. We will take this limit by writing $P_i = x_i P$, with $x_i$ fixed in the large-$P$ limit. One finds:
\begin{equation}\begin{aligned}\label{eq:C0HHL}
&\log C_0(P_0,x_1 P,x_2 P)\\
 =& \left(-4x_1^2\log(2x_1)-4x_2^2\log(2x_2)+2(x_1+x_2)^2\log(x_1+x_2)+2(x_1-x_2)^2\log(x_1-x_2)\right)P^2\\
 &-\pi Q(x_1+x_2)P + \left({2Q^2\over 3}+4P_0^2-{1\over 3}\right)\log P\\
 &+\log{2^{{1\over 6}(2Q^2-1)}(x_1x_2)^{{1\over 6}(Q^2+1)}(x_1^2-x_2^2)^{{1\over 6}(Q^2+12P_0^2-2)}\Gamma_0(b)^4\Gamma_b(2Q)\over\Gamma_b(Q)^3\Gamma_b(Q-2iP_0)\Gamma_b(Q+2iP_0)}+\mathcal{O}(P^{-1}).
\end{aligned}\end{equation}
The other interesting limit takes the difference $P_1-P_2 = 2\delta$ to be fixed, with the average $P\to\infty$. Note that in terms of dimensions $h$, this means that $h_1-h_2$ is of order $\sqrt{h}$. In this limit one finds the following asymptotics
\begin{equation}
\begin{aligned}\label{eq:C0HHLfixed}
	&\log C_0(P_0,P-\delta,P+\delta)\\
	=& -2\pi QP+2(h_0-4\delta^2)\log(P)\\
	&+\log{2^{{2Q^2-1-96\delta^2\over 6}}e^{-{Q^2\over 4}-3P_0^2-12\delta^2}\left(Q^2+4(P_0-2\delta)^2\right)^{{1\over 24}(Q^2+12(P_0-2\delta)^2)}\left(Q^2+4(P_0+2\delta)^2\right)^{{1\over 24}(Q^2+12(P_0+2\delta)^2)}\over(16P_0^4+8P_0^2(Q^2-16\delta^2)+(Q^2+16\delta^2)^2)^{1\over 12}}\\
	&+\log{\Gamma_0(b)^4\Gamma_b(2Q)\over\Gamma_b(Q)^3\Gamma_b(Q-2iP_0)\Gamma_b(Q+2iP_0)}+\mathcal{O}(P^{-1}).
\end{aligned}
\end{equation}
Several recent papers have studied asymptotics of the averaged off-diagonal heavy-heavy-light structure constants in CFT$_2$, including \cite{Brehm:2018ipf,Hikida:2018khg,Romero-Bermudez:2018dim}. The most directly comparable result is equation (2.33) of \cite{Brehm:2018ipf}, which studied these OPE asymptotics by considering the torus two-point function in a particular kinematic limit, imposing modular covariance, and performing an inverse Laplace transform to extract the spectral density. While the first line of our result (\ref{eq:C0HHLfixed}) reproduces the entropic suppression $e^{-S/2}$ expected from the eigenstate thermalization hypothesis, there appears to be a nontrivial difference between our subleading terms (written in terms of the dimensions and the central charge in equation (\ref{eq:HHL})) and those of \cite{Brehm:2018ipf}. Again, we would like to emphasize the technical simplicity of our argument, which does not rely on carefully establishing the behaviour of conformal blocks in simultaneous large-weight and kinematic limits.

\subsection{Genus-two partition function: heavy-heavy-heavy}

The final constraint from crossing we will study arises from modular invariance of the genus two partition function $G_{2,0}$. We will relate the conformal block decomposition in two channels, which we call `sunset' and `dumbbell'; these channels and the relation between them are illustrated in figure \ref{fig:genus2MS}.
\begin{equation}\begin{aligned}
	G_{2,0}
	&= \sum_{\mathcal{O}_1}\sum_{\mathcal{O}_2}\sum_{\mathcal{O}_3}C_{123}^2\mathcal{F}^{(\text{sunset})}(P_1,P_2,P_3)\bar{\mathcal{F}}^{(\text{sunset})}(\bar{P}_1,\bar{P}_2,\bar{P}_3)\\
	&= \int\left(\prod_{j=1}^3{dP_j\over 2}{d\bar P_j\over 2}\right)\rho_{\text{sunset}}(P_1,P_2,P_3,\bar P_1,\bar P_2, \bar P_3)\mathcal{F}^{(\text{sunset})}(P_1,P_2,P_3)\bar{\mathcal{F}}^{(\text{sunset})}(\bar{P}_1,\bar{P}_2,\bar{P}_3)\\
	&= \sum_{\mathcal{O}'_1}\sum_{\mathcal{O}'_2}\sum_{\mathcal{O}'_3}C_{1'1'2'}C_{2'3'3'}\mathcal{F}^{(\text{dumbbell})}(P'_1,P'_2,P'_3)\bar{\mathcal{F}}^{(\text{dumbbell})}(\bar{P}'_1,\bar{P}'_2,\bar{P}'_3)\\
	&= \int\left(\prod_{j=1}^3{dP_j\over 2}{d\bar P_j\over 2}\right)\rho_{\text{dumbbell}}(P_1,P_2,P_3,\bar P_1,\bar P_2, \bar P_3)\mathcal{F}^{(\text{dumbbell})}(P_1,P_2,P_3)\bar{\mathcal{F}}^{(\text{dumbbell})}(\bar{P}_1,\bar{P}_2,\bar{P}_3).
\end{aligned}\end{equation}
We have here suppressed the dependence of $G_{2,0}$ and the blocks on the moduli, since by now it is hopefully clear that we have no need of them. This is fortunate, because for $g\geq 2$ the description of the moduli spaces and relations between different channels becomes technically very challenging, and in particular, we must contend more directly with the factors arising from the conformal anomaly.

\begin{figure}
	\centering
	\includegraphics[width=.25\textwidth]{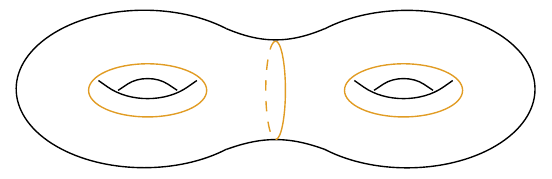} \raisebox{.03\textwidth}{\scalebox{1.5}{$=\int \frac{dP_1}{2}\frac{dP_2}{2}\modS_{P_1P_1'}[P'_3]\modS_{P_2P_2'}[P'_3]$}}\includegraphics[width=.25\textwidth]{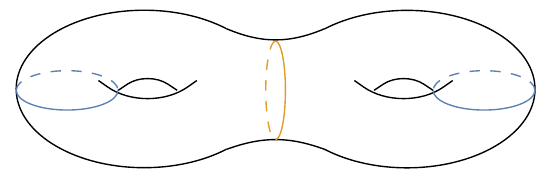} \\
		\flushright\raisebox{.03\textwidth}{\scalebox{1.5}{$=\int \frac{dP_1}{2}\frac{dP_2}{2}\frac{dP_3}{2}\modS_{P_1P_1'}[P'_3]\modS_{P_2P_2'}[P'_3] \fusion_{P_3P_3'}\sbmatrix{P_1&P_1\\P_2&P_2}$}}\includegraphics[width=.25\textwidth]{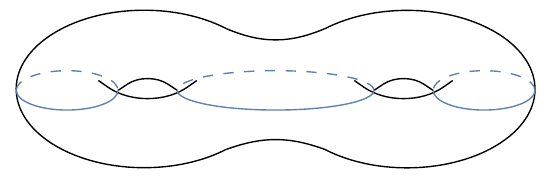}
	\caption{The sequence of moves expressing a genus 2 `dumbell' channel block in terms of `sunset' channel blocks.\label{fig:genus2MS}}
\end{figure}

To study the consequences of the genus-two modular crossing equation, we will employ the crossing kernel that relates dumbbell channel genus-two Virasoro blocks to those in the sunset channel. From figure \ref{fig:genus2MS}, we see that, like the crossing kernel for the torus two-point function, this kernel is simply a product of sphere four-point and torus one-point kernels:
\begin{equation}\begin{gathered}\label{eq:genusTwoKernel}
	\kernel_{P_1P_2P_3;P'_1P'_2P'_3} = \modS_{P_1P'_1}[P'_2]\modS_{P_3P'_3}[P'_2]\fusion_{P_2P'_2}\sbmatrix{ P_1 & P_3 \\ P_1 & P_3}\\
	\rho_\text{sunset}(P_i,\bar{P}_i) = \int \left(\prod_{i=1}^3 \frac{dP_i'}{2}\frac{d
	\bar{P}_i'}{2}\right) \kernel_{P_i;P'_i}\kernel_{\bar{P}_i;\bar{P}'_i}\,\rho_\text{dumbbell}(P'_i,\bar{P}'_i)
\end{gathered}\end{equation}

Once again, we will find that in appropriate limits, the spectral density in the sunset channel is dominated by the contribution of the identity in all internal cuffs of the dumbbell channel. The corresponding spectral density is given by the following identity kernel:
\begin{equation}\begin{aligned}\label{eq:genusTwoIdentity}
	\kernel_{P_1P_2P_3;\id\id\id} &=\modS_{P_1\id}[\id]\modS_{P_3\id}[\id] \fusion_{P_2\id}\sbmatrix{P_1 & P_3 \\ P_1 & P_3} \\
	&= \rho_0(P_1)\rho_0(P_2)\rho_0(P_3)C_0(P_1,P_2,P_3).
\end{aligned}\end{equation}
Thus, once again, the asymptotic behaviour of the OPE coefficients, now when all three operators are heavy, is determined by the asymptotics of the universal object $C_0(P_1,P_2,P_3)$. Precisely as in \eqref{eq:torusTwoPointId}, corrections to this asymptotic formula due to the exchange of non-vacuum primaries in the dumbbell channel are exponentially suppressed by the ratio
\begin{equation}
	{\kernel_{P_1 P_2 P_3;P'_1 P'_2 P'_3}\over\kernel_{P_1 P_2 P_3;\id\id\id}} \approx e^{-2\pi(\alpha_1'P_1+\alpha_3'P_3)}
\end{equation}
in the limit where the ratios or differences between the $P_i$ are held fixed. 
In the original dumbbell channel, $\alpha_2'$ cannot be nonzero unless both $\alpha_1'$ and $\alpha_3'$ are nonzero, so this is always exponentially small. More details are contained in appendix \ref{app:higherGenusAsymptoticCorrections}.

The conclusion is that the sunset channel OPE density is well-approximated by the exchange of the vacuum Verma module in the dumbbell channel when the internal weights all become heavy:
\begin{equation}
	\rho_{\text{sunset}}(P_1,\bar{P}_1;P_2,\bar{P}_2; P_3,\bar{P}_3) \approx \kernel_{P_1P_2P_3;\id\id\id}\kernel_{\bar{P}_1 \bar{P}_2 \bar{P}_3;\id\id\id},~P_i,\bar{P}_i\to \infty
\end{equation}
Thus the kernel (\ref{eq:genusTwoIdentity}) encodes an asymptotic formula for OPE coefficients in the heavy-heavy-heavy regime, averaged over the weights of all three heavy operators
\begin{equation}
	\overline{\left|C_{123}\right|^2}\sim C_0(P_1,P_2,P_3)C_0(\bar{P}_1,\bar{P}_2, \bar{P}_3),~P_i,\bar{P}_i\to\infty.
\end{equation}
As before, in the presence of a nonzero twist gap this formula holds at large spin in which only the left-moving momenta $P_1,P_2, P_3$ or the right-moving momenta $\bar{P}_1,\bar{P}_2, \bar{P}_3$ are taken to be large.

We can now recover asymptotic formulas for the microcanonical average of all heavy OPE coefficients from the relevant asymptotics of $C_0$. For example, if we fix ratios of $P_i$, parameterizing as $P_i = x_i P$ with $x_i>0$ fixed and $P\to\infty$, we have 
\begin{equation}\begin{aligned}\label{eq:C0HHH}
	&\log C_0(x_1 P, x_2 P, x_3 P)\\
	=& \left(-4\sum_{i=1}^3x_i^2\log(2x_i)+\sum_{\epsilon_2,\epsilon_3=\pm}(x_1+\epsilon_2 x_2+\epsilon_3 x_3)^2\log|x_1+\epsilon_2 x_2+\epsilon_3 x_3|\right)P^2\\
	&-\pi Q(x_1+x_2+x_3)P+\left({5Q^2-1\over 6}\right)\log P\\
	&+\log{2^{Q^2\over 2}(x_1x_2x_3)^{{1\over 6}(Q^2+1)}\prod_{\epsilon_2,\epsilon_3 = \pm}|x_1+\epsilon_2 x_2+\epsilon_3 x_3|^{{1\over 12}(Q^2-2)}\Gamma_0(b)^2\Gamma_b(2Q)\over\Gamma_b(Q)^3}+\mathcal{O}(P^{-1}).
\end{aligned}\end{equation}
In the case where $|P_i-P_j|$ is fixed in the limit, we instead have
\begin{equation}
\begin{aligned}\label{eq:C0HHHfixed}
	&\log C_0(P+\delta_1,P+\delta_2,P-\delta_1-\delta_2)\\
	=& 3P^2\log{27\over 16} -3\pi Q P +{1\over 6}(5Q^2-1)\log(P)\\
	&+\log{2^{{Q^2\over 2}-8(\delta_1^2+\delta_1\delta_2+\delta_2^2)}3^{{Q^2-2\over 12}}\Gamma_0(b)^2\Gamma_b(2Q)\over \Gamma_b(Q)^3}+\mathcal{O}(P^{-1}).
\end{aligned}
\end{equation}

These limits were studied using genus 2 modular invariance in \cite{Cardy:2017qhl}, using conformal block techniques. This analysis used the same underlying crossing relation, relating the heavy blocks in the sunset channel to the identity in the OPE channel (or, equivalently, a different necklace channel, obtained by an additional fusion move; the identity blocks in these two channels are identical). Results were only obtained for large $c$, where additional techniques to analyse conformal blocks are available, only included terms up to order $P\sim\sqrt{h}$ in $\log C_0$, and did not have a complete result for the term scaling exponentially in $P^2\sim h$ (the first line of \eqref{eq:C0HHHfixed}) valid at general ratios of operator dimensions. Nonetheless, all our formulas match those in \cite{Cardy:2017qhl}, including confirming a conjectured correction $c\to c-1$ from finite central charge. Our new method, with far less work, extends these results to higher orders and finite central charge.

\section{On the relation to Liouville theory}\label{sec:LiouvilleEquations}

In section \ref{sec:LiouvilleUniqueness}, we observed the relation between our universal object $C_0$ and the DOZZ formula for the structure constants of Liouville theory,
\begin{equation}\label{eq:DOZZ}
	C_0(P_1,P_2,P_3) \propto \frac{C_\mathrm{DOZZ}(P_1,P_2,P_3)} {\left(\prod_{k=1}^3S_0(P_k)\rho_0(P_k)\right)^{1\over 2}}.
\end{equation}
We then sketched an argument which explained this relation from a common origin in representation theory. We here give more details of that argument, explaining why the DOZZ formula must be constructed from the identity fusion kernel, as the unique solution to crossing built from only scalar Virasoro primaries.

To this end, we give general arguments for the identities which establish that the identity fusion kernel provides a solution to crossing with scalar primaries, applicable for any chiral algebra. Many of the methods are familiar in the context of rational CFTs. Secondly, we explicitly demonstrate that the relevant identities hold for the Virasoro crossing kernels of \cite{Ponsot:1999uf,Ponsot:2000mt,Teschner:2003at}, which is a consistency check that these arguments extend to this non-rational situation.

We perform this analysis for two cases. First, we study four-point crossing, where our arguments are very similar to those given in \cite{Ribault:2014hia}, for example. Secondly, we give similar arguments for modular S-invariance of the torus one-point function.

\subsection{Four-point crossing symmetry}

Following the general arguments of section \ref{sec:LiouvilleUniqueness}, the four-point crossing equation \eqref{eq:fusionTransform} with only scalar primaries becomes
\begin{equation}
	\fusion^{-1}_{P_tP_s}\, \rho_s(P_s) = \fusion_{P_sP_t} \, \rho_t(P_t),
\end{equation}
where we have inverted one of the fusion kernels to move it to the left hand side. We can write this relation with explicit dependence on the external operators as follows:
\begin{equation}\label{eq:fusionscalars}
	 \fusion_{t s} \sbmatrix{4 & 1 \\ 3 & 2} \rho_s =  \fusion_{s t} \sbmatrix{2 & 1 \\ 3 & 4 }\rho_t \;.
\end{equation}
We have used the fact that the inverse fusion kernel is the same as the fusion kernel with a permutation of external operators. Here and in the following, for brevity of notation we have suppressed momentum labels by replacing $P_i$ simply with $i$; in particular, the external operator labelled by $1$ is not to be confused with the identity representation, denoted by $\id$. Our aim in the following is to find an identity of the form \eqref{eq:fusionscalars}, and hence a solution to crossing. Note that if we have one solution to this equation, any other solution is related by multiplying $\rho_s$, $\rho_t$ by the same constant (independent of $P_s,P_t$, but not the external operators since we cannot fix their normalisations). The only exception to this occurs when the fusion kernel is block diagonal, in which case there is an independent solution for each block.
 
 To proceed, we make use of a consistency condition satisfied by fusion kernels, the famous pentagon identity, which in our notation reads
\begin{equation}\label{eq:pentagon}
	\sum_r \fusion_{r p} \sbmatrix{ 1 & q \\ 2 & 3 }  \fusion_{s q} \sbmatrix{1 & 5 \\ r & 4 } \fusion_{t r} \sbmatrix{2 & s \\ 3 & 4 } = \fusion_{t q} \sbmatrix{p& 5 \\ 3 & 4 } \fusion_{s p} \sbmatrix{1 & 5 \\ 2 & t }.
\end{equation}
We have written this with a sum over $r$, as appropriate for the fusion matrix in rational CFTs. For the $c>1$ Virasoro fusion kernels of \cite{Ponsot:1999uf,Ponsot:2000mt} with continuous families of representations, the sum becomes an integral with the appropriate measure. The identity follows from considering two possible sequences of fusion moves applied to the five-point conformal blocks, sketched in figure \ref{fig:pentagon}, which must act in the same way.
\begin{figure}[t]
	\centering
	\includegraphics[width=.7\textwidth]{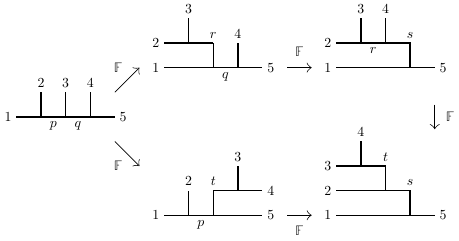}
	\caption{The sequence of crossing moves applied to the sphere five-point conformal block leading to the pentagon identity (\ref{eq:pentagon}) for the fusion kernel.}\label{fig:pentagon}
\end{figure}

We only require a special case of the identity, taking $q=\id$, which also sets $5=4$ and $p=3$ so that the blocks and fusion kernels are well-defined (otherwise, they become infinite signalling the disallowed fusion). The first fusion move then becomes trivial, giving a $\delta$-function that sets $r=1$
\begin{equation}
	\fusion_{r3}\sbmatrix{1& \id \\2&3} = \delta_{1r}.
\end{equation}
This relation can be explicitly verified for the Ponsot-Teschner fusion kernel
(\ref{eq:explicitFusion}) by taking the appropriate limit: the kernel vanishes at generic $P_r$ in the limit, but a delta-function $\delta(P_r-P_1)$ is produced by two poles which pinch the contour, with finite residue in the limit.
This leaves us with an identity without an internal sum,
\begin{equation}\label{eq:pentagon2}
	 \fusion_{s \id} \sbmatrix{1 & 4 \\ 1 & 4 } \fusion_{t 1} \sbmatrix{2 & s \\ 3 & 4 } = \fusion_{t \id} \sbmatrix{3 & 4 \\ 3 & 4 } \fusion_{s 3} \sbmatrix{1 & 4 \\ 2 & t },
\end{equation}
which one can check from the explicit form of the Ponsot-Teschner fusion kernel \eqref{eq:explicitFusion}. To see this, we note that we can rewrite the desired equality \eqref{eq:pentagon2} as 
\begin{equation}\label{eq:rewritePentagon2}
	\fusion_{s \id} \sbmatrix{1 & 4 \\ 1 & 4 } \fusion_{t \tilde 1} \sbmatrix{2 & s \\ 3 & \tilde 4 } = \fusion_{t \id} \sbmatrix{3 & 4 \\ 3 & 4 } \fusion_{s 3} \sbmatrix{2 & t\\1 & 4 },
\end{equation}
where by the tilded entries $\tilde i$, we mean that we replace $P_i\to -P_i$, an operation under which the fusion kernel is invariant. Written in this form, \eqref{eq:rewritePentagon2} is immediate from the expressions \eqref{eq:explicitFusion} after a shift of the variable in the contour integrals, and using $S_b(x) = S_b(Q-x)^{-1}$.

Now, by permuting labels in \eqref{eq:pentagon2} ($1\to t\to s\to 1$), we have
\begin{equation}\label{eq:pentagon3}
	 \fusion_{1 \id} \sbmatrix{t & 4 \\ t & 4 } \fusion_{s t} \sbmatrix{2 & 1 \\ 3 & 4 } = \fusion_{s \id} \sbmatrix{3 & 4 \\ 3 & 4 } \fusion_{1 3} \sbmatrix{t & 4 \\ 2 & s },
\end{equation}
where we recognise one term on the left as the fusion kernel of interest in \eqref{eq:fusionscalars}. By another permutation of labels, swapping $2\leftrightarrow 4$ and $t\leftrightarrow s$ in \eqref{eq:pentagon3}, we find an identity involving the inverse fusion kernel of interest,
\begin{equation}\label{eq:pentagon4}
	 \fusion_{1 \id} \sbmatrix{s & 2 \\ s & 2 } \fusion_{t s} \sbmatrix{4 & 1 \\ 3 & 2} = \fusion_{t \id} \sbmatrix{3 & 2 \\ 3 & 2 } \fusion_{1 3} \sbmatrix{s & 2 \\ 4 & t }.
\end{equation}
Now, since the fusion kernels are invariant under exchange of rows or columns, the $\fusion_{1 3}$ kernels appearing in the two identities are the same, so we can combine \eqref{eq:pentagon3} and \eqref{eq:pentagon4} to find
\begin{equation}\label{eq:pentagon5}
	\fusion_{s \id} \sbmatrix{3 & 4 \\ 3 & 4 }\fusion_{1 \id} \sbmatrix{s & 2 \\ s & 2 } \fusion_{t s} \sbmatrix{4 & 1 \\ 3 & 2} =  \fusion_{s t} \sbmatrix{2 & 1 \\ 3 & 4 } \fusion_{1 \id} \sbmatrix{t & 4 \\ t & 4 }\fusion_{t \id} \sbmatrix{3 & 2 \\ 3 & 2 }\, .
\end{equation}
This is an identity precisely of the form \eqref{eq:fusionscalars} and hence a scalar solution to four-point crossing, with
\begin{equation}
	\rho_s = k\, \fusion_{s \id} \sbmatrix{3 & 4 \\ 3 & 4 }\fusion_{1 \id} \sbmatrix{s & 2 \\ s & 2 },\quad \rho_t = k\, \fusion_{1 \id} \sbmatrix{t & 4 \\ t & 4 }\fusion_{t \id} \sbmatrix{3 & 2 \\ 3 & 2 }\, ,
\end{equation}
where $k$ is independent of $P_s,P_t$, but otherwise arbitrary. Using the expression \eqref{eq:idFusion} for the identity fusion kernel in terms of our universal functions $C_0$ and $\rho_0$, we can write this solution as
\begin{equation}
	\rho_s(P_s) = k\, \rho_0(P_s)C_0(P_1,P_2,P_s)C_0(P_s,P_3,P_4),\quad \rho_t = k\, \rho_0(P_t)C_0(P_1,P_4,P_t)C_0(P_t,P_2,P_3)\, ,
\end{equation}
where a factor of $\rho_0(P_1)$ has been absorbed into $k$. From the relation \eqref{DOZZ} between $C_0$ and the DOZZ formula, we see that $\rho_s$ and $\rho_t$ are precisely the S- and T-channel spectral densities in Liouville theory, making an appropriate choice of $k$.

\subsection{S-invariance of torus one-point functions}

We now make a similar argument to show that Liouville theory provides the unique modular covariant torus one-point functions for scalars. We begin by writing the equation for one-point S-invariance \eqref{eq:modularScov} for scalars in a form analogous to \eqref{eq:fusionscalars}, using the fact that $\modS$ is its own inverse:
\begin{equation}\label{eq:Sscalars}
	\modS_{ts}[\op]\rho_\op(s) = \modS_{st}[\op]\rho_\op(t).
\end{equation}
Here, the same torus one-point spectral density $\rho_\op$ (where $\rho_\op(s)$ is the density of internal states $\rho(s)$ times OPE coefficients $C_{\op ss}$) appears on both sides.

To find a relation of the form \eqref{eq:Sscalars}, we require an identity for the modular S-kernel $\modS$ to play an analagous role to the pentagon identity in the above. Such an identity arises from consistency of torus two-point functions, where two different sequences of moves applied to a vacuum block must be equivalent:
\begin{equation}
	\fusion_{\op\id}\sbmatrix{t&t\\t&t} \modS_{st}[\op] = \modS_{s\id}[\id] \sum_u e^{2\pi i(h_s+h_t-h_u-h_\op/2)} \fusion_{u\id}\sbmatrix{s&t\\s&t} \fusion_{\op u}\sbmatrix{t&t\\s&s}.
\end{equation}
This identity is well-known for rational theories \cite{Moore:1988uz,Moore:1989vd}, but also applies to the Virasoro kernels at generic central charge \cite{Teschner:2008qh}, with the sum over $u$ replaced by an integral with appropriate measure. For rational theories, this identity is the key to proofs of the Verlinde formula \cite{Moore:1988uz,Moore:1989vd}, so these considerations can be applied to explore analogues of the Verlinde formula for irrational theories \cite{Teschner:2008qh}.



For us, the most important feature of this identity is that the right-hand side is symmetric under swapping $s\leftrightarrow t$, except for the factor of the identity S-matrix $\modS_{s\id}[\id]$. From this observation, we find the simple relation
\begin{equation}
	 \modS_{st}[\op] \modS_{t\id}[\id]\fusion_{\op\id}\sbmatrix{t&t\\t&t} = \modS_{ts}[\op] \modS_{s\id}[\id]\fusion_{\op\id}\sbmatrix{s&s\\s&s}.
\end{equation}
This identity is precisely of the desired form \eqref{eq:Sscalars}, with
\begin{equation}
	\rho_\op(P) \propto \modS_{P\id}[\id]\fusion_{\op\id}\sbmatrix{P&P\\P&P} = \rho_0(P_\op)\rho_0(P)C_0(P_\op,P,P).
\end{equation}
Up to a $P$-independent normalisation constant, this is precisely the torus one-point spectral density for Liouville theory constructed from the DOZZ formula \eqref{DOZZ}.


\section{Semiclassical limits}\label{sec:semiclassical}
Throughout this paper we have emphasized that our asymptotic formulas apply in any two-dimensional irrational CFT for any $c>1$, providing universal results in a kinematic limit of large dimension or spin. However, it is natural to expect our results to be particularly powerful in holographic theories with a weakly coupled AdS$_3$ dual, and to have a corresponding gravitational interpretation. The basic reason for this is simple: the corrections to the asymptotic formula come from the lightest operators in the theory, and existence of a holographic dual requires having few such operators (a sparse light spectrum) \cite{Heemskerk:2009pn,ElShowk:2011ag,Hartman:2014oaa}. For example, in higher dimensions generic theories contain double-twist operators with anomalous dimensions suppressed at large spin \cite{Komargodski:2012ek,Fitzpatrick:2012yx}; in holographic theories, the `t Hooft limit extends this to double-trace operators with anomalous dimensions suppressed at large $N$, now at finite spin. The corresponding gravitational interpretation involves two-particle states in AdS, which generically are weakly interacting only with very large orbital angular momentum, when the particles are widely separated, but in holographic theories also interact weakly at finite separation. An example in $d=2$ is the density of states, which for holographic theories is given by the Cardy formula not just for very heavy operators, but also at large $c$ for energies of order $c$ \cite{Hartman:2014oaa}, interpreted as the Bekenstein-Hawking entropy of BTZ black holes \cite{Strominger:1997eq}.

With this in mind, in this section we will give gravitational interpretations of our universal OPE coefficients $C_0$ in various large $c$ limits. We will not attempt here to pin down precisely when these formulas apply, in terms of constraints on the theory and regime of operator dimensions; see \cite{Michel:2019vnk} for recent work in this direction.

Nonetheless, it is simpler to interpret and understand this regime in the gravitational description. Since our formulas come from expanding an identity block in an alternative channel, we can interpret our formulas as a microcanonical version of `vacuum block domination', giving the density of states in a regime where a correlation function is well-approximated by only the identity Virasoro block in the appropriate channel 
\cite{Hartman:2013mia,Roberts:2014ifa,Anous:2016kss,Kraus:2018pax}. At large $c$, an identity block is given by the gravitational action of a particular locally empty AdS solution (which could be a BTZ black hole or handlebody at higher genus), along with worldlines of particles propagating between external operator insertions \cite{Faulkner:2013ica,Hijano:2015rla,Kraus:2017ezw,Dong:2018esp,Maxfield:2017rkn}:
\begin{equation}
	\mathcal{F}_\id \approx e^{- c\, S_\text{grav}}
\end{equation}
We therefore expect our formulas to be applicable when the gravitational path integral is dominated by such a solution, up to loop corrections\footnote{Note that  the identity block itself need not be a larger contribution than any other block. Corrections at one-loop order change the coefficient of $e^{- c\, S_\text{grav}}$, and come from light operator exchanges of the same order as $\mathcal{F}_\id$.}. This holds for a kinematic regime of parametrically low temperature or small cross-ratios, but for holographic theories is expected to extend to a regime of kinematics which are fixed in the large $c$ limit. The question is how far this regime extends before encountering a phase transition. The simplest such phase transitions are first-order `Hawking-Page' transitions, where an identity block in different channel dominates. However, note that even for local, weakly coupled gravitational theories, there need not be any channel in which the vacuum dominates: for example, there may be a phase in which a scalar field condenses after a second-order phase transition \cite{Belin:2017nze,Dong:2018esp}. Vacuum dominance potentially particularly subtle for correlation functions in kinematic regimes such as those with operators out of time order \cite{Chang:2018nzm}.

We now give our examples of gravitational interpretations of the universal OPE coefficients $C_0$ in various limits. These are all explored in more detail elsewhere, but we present them here together as consequences of the same formula, emphasising the unifying nature of our results. Furthermore, the list may well not be exhaustive, since we have not included all possible semiclassical limits, and our understanding of the connections to gravity is far from complete.

\subsection{Spectral density of BTZ black holes}
 
For our first example, we take a large $c$ limit of $C_0$ which probes the physics of BTZ black holes. We take two operators to be heavy, with dimensions $h_1,h_2$ scaling with $c$, to correspond to black hole states, but with similar dimensions, $h_1-h_2$ fixed as $c\to\infty$. The third operator, acting as a probe of the geometry, has $h$ fixed in the limit. In terms of the momentum variables $P$, we take
 \begin{equation}
 P_1= b^{-1} p + b\delta \, \quad  P_2=b^{-1}p-b\delta, \quad P_3 = i\left(\tfrac{Q}{2}-b h\right),
 \end{equation}
 and fix $p,\delta,h$ in the $b\to 0$ limit. We can then interpret $C_0$ as governing the matrix elements $\langle \text{BH}_2|\op|\text{BH}_1\rangle$ of the probe operator $\op$ of dimension $h$ between black hole states of nearby energies.

This limit of the fusion kernel was studied in \cite{Collier:2018exn}, with the result
 \begin{equation}
 \rho_0(b^{-1} p) C_0\left(P_1,P_2,P_3\right)\sim \frac{(2p)^{2h}}{ 2\pi b}\frac{\Gamma(h+2i\delta)\Gamma(h-2i\delta)}{\Gamma(2h)}.
 \end{equation}
This is the left-moving half spectral density associated to free matter propagating an a BTZ black hole background\footnote{Strictly speaking, this holds in a case where we are insensitive to the compactness of the spatial circle, either large black holes or heavy external operators.} 
 \cite{Birmingham:2001pj}. In particular, the poles at imaginary $\delta$ are associated with the frequencies of quasinormal modes governing the approach to equilibrium. This result is sufficient to recover the `heavy-light' limit of conformal blocks \cite{Fitzpatrick:2014vua,Fitzpatrick:2015zha}; see \cite{Collier:2018exn} for more details.

\subsection{Near-extremal BTZ and the Schwarzian theory}

Our second example (based on results to appear \cite{Ghosh:2019rcj}) is similar to the first, but treats the distinct case where the black hole of interest is very close to extremality.

Rotating BTZ black holes exist for dimensions above the extremality bound $h>\frac{c-1}{24}$, and we will tune our operators close to this, with $h-\frac{c-1}{24}$ of order $c^{-1}$. Our third operator will remain a light probe. This means we have
 \begin{equation}
 P_1= b k_1  \, \quad  P_2=b k_2, \quad P_3 = i\left(\tfrac{Q}{2}-b h\right),
 \end{equation}
 where we fix $k_1,k_2,h$ and take $b\to 0$.

In this limit, our universal density of states $\rho_0$ and OPE coefficients $C_0$ are given by
\begin{align}
	\rho_0(b k) &\sim 8\sqrt{2} \pi b^2 k \sinh(2\pi k) \label{eq:rho0limit} \\
	C_0\left(b k_1,b k_2,i(\tfrac{Q}{2}-bh)\right) &\sim  \frac{b^{4h}}{\sqrt{2}(2\pi b)^3} \frac{\prod_{\pm\pm}\Gamma(h\pm ik_1\pm ik_2)}{\Gamma(2h)},
\end{align}
where the $\prod_\pm$ refers to a product of four terms with all possible sign combinations. These expression may be familiar from the Schwarzian theory, which governs the dynamics of weakly broken conformal symmetry \cite{Maldacena:2016hyu,Maldacena:2016upp,Mertens:2017mtv}. This theory arises in near-extremal black holes, which have a near-horizon AdS$_2$ region with dynamics governed by Jackiw-Teitelboim gravity \cite{Almheiri:2014cka,Maldacena:2016upp}. Specifically, $\rho_0$ is proportional to the density of states for the Schwarzian theory, and $C_0$ to a transition amplitude appearing in calculations of correlation functions \cite{Mertens:2017mtv, Kitaev:2018wpr, Yang:2018gdb}.

The appearance of these quantities is a sign that there is a universal sector of large $c$ CFTs which knows about quantum geometry, where the metric fluctuations are not suppressed. The connection between the Schwarzian theory, near-extremal BTZ and universality in CFT will be explored in much greater detail in forthcoming work \cite{Ghosh:2019rcj}.

\subsection{Conical defect action}

Finally, we consider a regime where all three operators have dimensions scaling with $c$. If we take $\frac{24h}{c}>1$ in this limit, as required for asymptotic formulas, $C_0$ should be interpreted as giving a three-point function of black hole microstates. It is unclear whether there is a direct calculation of this quantity, giving the semiclassical limit of $C_0$ as an on-shell action. However, perhaps surprisingly, if we fix $\frac{24h}{c}<1$ and take $c\to\infty$, there is such an interpretation, shown in \cite{Chang:2016ftb}. Those authors computed the vacuum fusion kernel in a large central charge limit,
\begin{equation}
	\alpha_i = b^{-1}\eta_i, \quad b\to 0,\quad \text{ fixed }\eta_i,\quad i=1,2,3,
\end{equation}
and equated it to a suitably regularised on-shell action of a geometry corresponding to three heavy particles running between the asymptotic boundary and a trivalent vertex. The action in this case is Einstein-Hilbert, plus an action $m_i L_i$ for each particle, where $L_i$ is a regularised proper length of the particle's worldline and $m_i\sim \frac{c}{3}\eta_i$ is its mass. Since the particles have masses of order $c$, they backreact to form three conical defects in the geometry, meeting at the vertex\footnote{No particle action was included in \cite{Chang:2016ftb}, but they also included no singular contribution to the Einstein Hilbert action localised on the worldline. These two terms are equal and opposite, so the results are equivalent.}.

In our notation, we can express the result of \cite{Chang:2016ftb} as a limit of $C_0$:
\begin{equation}
	\begin{gathered}
		\log C_0 \sim b^{-2}\Big(-\tfrac{1}{2} S_\text{grav}(\eta_1,\eta_2,\eta_3)+ i \theta(\eta_1,\eta_2,\eta_3)\Big), \\
		\begin{aligned}
	-\tfrac{1}{2} S_\text{grav} &= \left(F(2\eta_1)-F(\eta_2+\eta_3-\eta_1)+(1-2\eta_1)\log(1-2\eta_1)+\text{(2 permutations)}\right)\\
		&\quad + F(0) -F(\eta_1+\eta_2+\eta_3)-2(1-\eta_1-\eta_2-\eta_3)\log(1-\eta_1-\eta_2-\eta_3)\\
		\theta &= \pi(\eta_1+\eta_2+\eta_3-1),
		\end{aligned}
	\end{gathered}
\end{equation}
where $F(z) = I(z) + I(1-z)$ for  $I(z) = \int_\half^z dy\log\Gamma(y)$. The action $b^{-2}S_\text{grav}$ appearing here is precisely the gravitational action for the conical defect network described above. When left- and right-moving sectors are combined, for scalars the phase $\theta$ cancels.

When conformal blocks are computed at large $c$ as an on-shell gravitational action, this conical defect action, and hence this limit of $C_0$, appear as the natural normalisation of the blocks \cite{Maloney:2016kee,Dong:2018esp}. While the relation with our universal asymptotic formulas is suggestive, it remains rather mysterious from that point of view, and deserves to be better understood.

\section{Torus one-point functions \& the Eigenstate Thermalization Hypothesis}\label{sec:ETH}

Although the primary focus of our paper is on the asymptotic behaviour of the $C_{ijk}{}^2$, similar techniques can be applied to other observables in two-dimensional conformal field theory.  For example, by studying the modular covariance of the torus one-point function of an operator $\mathcal{O}_0$ one obtains an asymptotic formula for diagonal heavy-heavy-light structure constants $\overline{C_{OHH}}$, where we average over the heavy operator $H$. 
This was discussed in \cite{Kraus:2016nwo}, who found
\begin{equation}\label{km}
	\overline{C_{0HH}} \approx \mathcal{N}_0C_{0\chi\chi}\left(\Delta_H- {c-1\over 12}\right)^{\Delta_0/2}\exp\left[-{\pi(c-1)\over 3}\left(1-\sqrt{1-{12\Delta_\chi\over c-1}}\right)\sqrt{{12\Delta_H\over c-1}-1}\right],
\end{equation}
in the limit that $\Delta_H\to\infty$. Here $\chi$ is the lightest operator to which $\calo_0$ couples (i.e. for which $C_{0\chi\chi}\ne 0$), and is assumed to be sufficiently light, $\Delta_\chi < {c-1\over 12}$. The normalization factor $\mathcal{N}_0$ depends only on $c,\,\Delta_\chi$ and $\Delta_0$.  This analysis was performed at the level of the scaling blocks in \cite{Kraus:2016nwo} and was generalized to include the contribution of global blocks in \cite{Kraus:2017ezw}.  When regarded as a formula for the average value of the primary operators, however, equation (\ref{km}) is true only at leading order in $1/c$; the inclusion of Virasoro blocks provides corrections which are only subleading at large $c$.

We can now write down the finite $c$ version of this formula using the modular S kernel introduced in section \ref{sec:modularS} for torus one point functions. Following the same logic that led to our other asymptotic formulas, we conclude that
\begin{equation}
	\overline{C_{0HH}} \approx C_{0\chi\chi}{\modS_{P_HP_\chi}[P_0]\modS_{\bar P_H\bar P_\chi}[\bar P_0]\over \rho_0(P_H)\rho_0(\bar P_H)},~P_H,\bar P_H\to\infty
\end{equation}
provided that $\chi$, the lightest operator that couples to $\mathcal{O}_0$, is sufficiently light ($\alpha_\chi$ lies in the discrete range in the sense of \cite{Collier:2018exn}) and that there exists a gap above this lightest operator so that corrections due to the inversion of the contributions of other operators in the original channel are indeed suppressed. 
The large $P$ asymptotics of this formula are straightforward to find by taking the large $P_H$ limit of the modular S kernel, namely
\begin{equation}\label{eq:modSOverRho0}
	{\modS_{P_HP_\chi}[P_0]\over\rho_0(P_H)}\approx e^{-4\pi\alpha_\chi P_H}P_H^{h_0}.
\end{equation}
This reproduces the earlier result \eqref{km} in the appropriate limit.

We would like to emphasize two important qualitative differences between this formula and our other asymptotic formulas. The first is that it is not universal in the same sense as our other formulas, as it explicitly depends on the lightest operator that couples to $\mathcal{O}_0$, both through its conformal weights and OPE coefficient (this is because the vacuum Verma module cannot propagate as an intermediate state in either channel of the torus one-point function). Second, its derivation is on even less rigorous footing than our other asymptotic formulas because the structure constants that appear in the conformal block decomposition of the torus one-point function need not be positive, and so the spectral densities $\rho[\mathcal{O}_0],\,\tilde\rho[\mathcal{O}_0]$ do not in general have definite sign and may oscillate when integrated. This is unlike the product of structure constants that appear in the necklace channel conformal block decomposition of the torus two-point function of identical operators or the sunset channel of the genus-two partition function, which are positive in a unitary CFT. In fact, if the lightest operator that couples to $\mathcal{O}$ is sufficiently heavy (in particular, if it has twist $>{c-1\over 12}$), then one cannot even argue that the asymptotics of the structure constants are universal as corrections due to the propagation of other operators in the original channel are not parametrically suppressed.

As discussed in section \ref{sec:introETH}, the fact that the averaged diagonal heavy-heavy-light OPE coefficients are exponentially suppressed (via e.g. (\ref{eq:modSOverRho0})) implies a different hierarchy of suppression between the averaged diagonal and non-diagonal heavy-heavy-light structure constants than would naively have been expected from the usual statement of the Eigenstate Thermalization Hypothesis, where $f^{\calo}$ is order one and $g^{\calo}\approx e^{-\half S(\Delta)}$. Indeed, if the lightest operator that couples to $\mathcal{O}_0$ satisfies $\text{Re}(\alpha_\chi+\bar\alpha_\chi) \ge {Q\over 2}$ (for scalars, this corresponds to dimension $\Delta_\chi\ge {c-1\over 16}$), then there is \textit{no} suppression whatsoever of the averaged off-diagonal structure constants compared to the diagonal, and indeed the diagonal terms may be even smaller than the off-diagonal in this regime. This may be seen by comparing equation (\ref{eq:modSOverRho0}) with equation (\ref{eq:C0HHLfixed}). This contrast is particularly sharp in holographic theories with a large gap in the spectrum of primary operators, with only Planckian degrees of freedom. Indeed the dual of a theory of ``pure'' quantum gravity in $AdS_3$ is in a sense one where the averaged diagonal heavy-heavy-light structure constants are smallest.

\section*{Acknowledgements}
We would like to thank M.~Aganagic, N.~Benjamin, J.~Cardy, S.~Caron-Huot, S.~Rychkov, P.~Saad, G.J.~Turiaci, H.~Verlinde and X.~Yin for useful discussions.
A.M.~acknowledges the support of the Natural Sciences and Engineering Research Council of Canada (NSERC), funding reference number SAPIN/00032-2015. H.M.~is supported by a Len DeBenedictis Postdoctoral Fellowship and NSF grant PHY1801805, and receives additional funds from the University of California. I.T.~is supported by the Alexander S. Onassis Foundation under the contract F ZM 086-1. S.C. and A.M. thank the organizers of the 2019 meeting of the Simons collaboration on the non-perturbative bootstrap at Perimeter Institute, where some of this work was done. Research at Perimeter Institute is supported in part by the Government of Canada through the Department of Innovation, Science and Economic Development Canada and by the Province of Ontario through the Ministry of Economic Development, Job Creation and Trade. This work was supported in part by a grant from the Simons Foundation (385602, A.M.).
\appendix

\section{Explicit forms of elementary crossing kernels}\label{app:explicitForms}
In this section we will review the explicit forms of the elementary crossing kernels used in this paper, with a focus on the analytic structure of the kernels as a function of the intermediate weights. 

\subsection{Sphere four-point}
We will start by reviewing the explicit form of the fusion kernel, which implements the fusion transformation relating sphere four-point Virasoro conformal blocks in different OPE channels (see equation (\ref{eq:fusionTransformation})). The fusion kernel was worked out in explicit detail by Ponsot and Teschner \cite{Ponsot:1999uf,Ponsot:2000mt}. The expression involves the special functions $\Gamma_b(x)$, which is a meromorphic function with no zeros that one may think of as a generalization of the ordinary gamma function, but with simple poles at $x = -(mb+nb^{-1})$ for $m,n\in\mathbb{Z}_{\ge 0}$, and 
\begin{equation}
	S_b(x) = {\Gamma_b(x)\over \Gamma_b(Q-x)}.
\end{equation}
Many properties of these special functions, including large argument and small $b$ asymptotics, were summarized in \cite{Collier:2018exn} (see in particular appendix A of that paper). The explicit expression for the kernel involves a contour integral and is given by
\begin{equation}\label{eq:explicitFusion}
	\fusion_{P_sP_t}\sbmatrix{ P_2 & P_1 \\ P_3 & P_4 } = P_b(P_i;P_s,P_t)P_b(P_i;-P_s,-P_t)\int_{\mathcal{C}'}{ds\over i}\prod_{k=1}^4{S_b(s+U_k)\over S_b(s+V_k)},
\end{equation}
where the prefactor $P_b$ is given by
\begin{equation}
\begin{aligned}\label{eq:fusionPrefactor}
	&P_b(P_i;P_s,P_t)\\
	=& {\Gamma_b({Q\over 2}+i(P_s+P_3-P_4))\Gamma_b({Q\over 2}+i(P_s-P_3-P_4))\Gamma_b({Q\over 2}+i(P_s+P_2-P_1))\Gamma_b({Q\over 2}+i(P_s+P_1+P_2))\over \Gamma_b({Q\over 2}+i(P_t+P_1-P_4))\Gamma_b({Q\over 2}+i(P_t-P_1-P_4))\Gamma_b({Q\over 2}+i(P_t+P_2-P_3))\Gamma_b({Q\over 2}+i(P_t+P_2+P_3))}{\Gamma_b(Q+2iP_t)\over \Gamma_b(2iP_s)}
\end{aligned}
\end{equation}
and the arguments of the special functions in the integrand are
\begin{equation}
	\begin{split}
		U_1&=i(P_1-P_4)\\
		U_2&=-i(P_1+P_4) \\
		U_3&= i(P_2+P_3)\\
		U_4&=i(P_2-P_3)
	\end{split}
	\qquad
	\begin{split}
		V_1 &= Q/2+i(-P_s+P_2-P_4)\\
		V_2 &= Q/2+i(P_s+P_2-P_4) \\
		V_3 &= Q/2+iP_t \\
		V_4 &= Q/2-iP_t
	\end{split}
\end{equation}
The contour $\mathcal{C}'$ runs from $-i\infty$ to $i\infty$, traversing between the towers of poles running to the left at $s = -U_i-mb-nb^{-1}$ and to the right at $s = Q-V_j +mb+nb^{-1}$ in the complex $s$ plane, for $m,n\in\mathbb{Z}_{\ge0}$.

Viewed as a function of the internal weight $P_s$, the kernel (\ref{eq:explicitFusion}) has eight semi-infinite lines of poles extending to both the top and bottom of the complex plane
\begin{equation}
\begin{aligned}\label{eq:fusionPoles}
	\fusion_{P_sP_t}\sbmatrix{ P_2 & P_1 \\ P_3 & P_4 }\text{: }&\text{simple poles at }P_s = \pm i\left({Q\over 2}+iP_0+mb+nb^{-1}\right),\text{ for }m,n\in\mathbb{Z}_{\ge 0},\\
	&\text{where }P_0 = P_1+P_2,\, P_3+P_4 \text{ (and six permutations under reflection $P_i\to-P_i$)}.
\end{aligned}
\end{equation}
Roughly, half of these poles are explicit singularities of special functions in the prefactor (\ref{eq:fusionPrefactor}), while the other half arise from singularities of the contour integral, which occur when poles of the integrand pinch the contour. In the case particularly relevant for this paper of pairwise identical operators $P_4 = P_1,~P_3 =P_2$, these singularities are enhanced to double poles, although there is an exception when the T-channel internal weight $P_t$ is degenerate ($P_t = \pm {i\over 2} ((m+1)b+(n+1)b^{-1}),~m,n\in\mathbb{Z}_{\ge 0}$), in which case the poles remain simple when the external operators have weights consistent with the fusion rules.
 
In most cases, the contour of integration over the internal weight $P_s$ in the fusion transformation (\ref{eq:fusionTransformation}) can be taken to run along the real axis. However, as emphasized in \cite{Collier:2018exn,Kusuki:2018wpa}, when the external operators are sufficiently light, in particular when
\begin{equation}
	\text{Re}(i(P_1+P_2)) < -{Q\over 2}\text{ or }\text{Re}(i(P_3+P_4))<-{Q\over 2}
\end{equation}
then some poles of the fusion kernel (\ref{eq:fusionPoles}) cross the real $P_s$ axis and the contour must be deformed, leading to a finite number of discrete residue contributions to the S-channel decomposition of the T-channel Virasoro block. These correspond to the Virasoro analog of double-twist operators \cite{Collier:2018exn}.

In the special case of pairwise identical operators with T-channel exchange of the identity, the contour integral can be computed very explicitly and the fusion kernel takes the following simple form, which makes the analytic structure manifest
\begin{equation}\begin{aligned}\label{eq:vacFusion}
	\fusion_{P_s\id}\sbmatrix{ P_2 & P_1 \\ P_2 & P_1 } &= {\Gamma_b(2Q)\over \Gamma_b(Q)^3}{\Gamma_b({Q\over 2}+i(P_1+P_2-P_s))\times(\text{7 permutations under reflection $P\to -P$})\over \Gamma_b(2iP_s)\Gamma_b(-2iP_s)\Gamma_b(Q+2iP_1)\Gamma_b(Q-2iP_1)\Gamma_b(Q+2iP_2)\Gamma_b(Q-2iP_2)}\\
	&= \rho_0(P_s)C_0(P_1,P_2,P_s).
\end{aligned}\end{equation}

\subsection{Torus one-point}
The crossing kernel that implements the modular S transformation on torus one-point Virasoro blocks (see equation (\ref{eq:modularSTransform})) was worked out by Teschner \cite{Teschner:2003at}. Similarly to the fusion kernel, its explicit form involves a contour integral and is given by
\begin{equation}\begin{aligned}\label{eq:explicitModularS}
	\modS_{PP'}[P_0] =& {\rho_0(P)\over S_b({Q\over 2}+iP_0)}{\Gamma_b(Q+2iP')\Gamma_b(Q-2iP')\Gamma_b({Q\over 2}+i(2P-P_0))\Gamma_b({Q\over 2}-i(2P+P_0))\over\Gamma_b(Q+2iP)\Gamma_b(Q-2iP)\Gamma_b({Q\over 2}+i(2P'-P_0))\Gamma_b({Q\over 2}-i(2P'+P_0))}\\
	&\int_C{d\xi\over i}e^{-4\pi P'\xi}{S_b(\xi+{Q\over 4}+i(P+\half P_0))S_b(\xi+{Q\over 4}-i(P-\half P_0))\over S_b(\xi+{3Q\over 4}+i(P-\half P_0))S_b(\xi+{3Q\over 4}-i(P+\half P_0))}\\
	\equiv& Q_b(P,P',P_0)\int_C{d\xi\over i}e^{-4\pi P'\xi}T_b(\xi,P,P_0).
\end{aligned}\end{equation}
This integral representation only converges when 
\begin{equation}\label{eq:alphaPrimeCondition}
	 {1\over 2}{\rm Re}(\alpha_0)<{\rm Re}(\alpha')<{\rm Re}\left(Q-{1\over 2}\alpha_0\right).
\end{equation}
Outside of this range, the kernel is defined via analytic continuation, using the fact that it satisfies a shift relation that we will make explicit shortly.

The integral contributes the following series of poles in the $P$ plane, one extending to the top and the other extending to the bottom
\begin{equation}\begin{aligned}
	\text{integral: poles at }P = \pm{i\over 2}\left({Q\over 2}+iP_0+mb+nb^{-1}\right),~m,n\in\mathbb{Z}_{\ge 0}.
\end{aligned}\end{equation}
Together with the prefactor, the full kernel has the following polar structure in the $P$ plane
\begin{equation}\begin{aligned}
	\modS_{PP'}[P_0]:\text{ poles at }P = {i\over 2}\left({Q\over 2}-iP_0+mb+nb^{-1}\right),~m,n\in\mathbb{Z}_{\ge 0}, \text{ and all possible reflections (in $P,P_0$)}.
\end{aligned}\end{equation}
One can think of these poles as arising in the case that the external operator is a (Virasoro) double-twist of the internal operator.
Unlike the case of the fusion kernel, for unitary values of the weights none of these poles can cross the contour of integration $\text{Im}(P) = 0$.

Similarly to the case of the fusion kernel, the modular S kernel can be straightforwardly evaluated in the case that the external operator is the identity, $P_0 = i{Q\over 2}$. In this case, the prefactor vanishes and so we only need to extract the singularities of the contour integral. By carefully studying this limit, one finds
\begin{equation}
	\modS_{PP'}[\id] = 2\sqrt{2}\cos(4\pi P P'),
\end{equation}
precisely reproducing the non-degenerate modular S matrix for the Virasoro characters (\ref{eq:modularS}). To study the limit in which the internal operator in the original channel is also the identity one must be more careful, for the simple reason that the Virasoro vacuum character is not the same as the $h'\to 0$ limit of the non-degenerate Virasoro character; in the latter case, there are null states that do not decouple continuously. 

To study this limit more carefully, we note that the modular kernel satisfies the following shift relation (see e.g. \cite{Nemkov:2015zha})
\begin{equation}
\begin{aligned}\label{eq:modSShift}
	2\cosh(2\pi b P)\modS_{PP'}[P_0]=&\bigg({\Gamma(b(Q+2iP'))\Gamma(2ibP')\over\Gamma(b({Q\over 2}+i(2P'-P_0)))\Gamma(b({Q\over 2}+i(2P'+P_0)))}\modS_{P,P'-i{b\over 2}}[P_0]\\
	&+{\Gamma(b(Q-2iP'))\Gamma(-2ibP')\over \Gamma(b({Q\over 2}-i(2P'+P_0)))\Gamma(b({Q\over 2}-i(2P'-P_0)))}\modS_{P,P'+i{b\over 2}}[P_0]\bigg).
\end{aligned}
\end{equation}
Now consider the limit $P'\to i{b^{-1}\over 2}$ of this equality. The first term on the right-hand side will be singular unless we take $P_0$ to $i{Q\over 2}$ at the same time. To facilitate the study of this limit, we write $P' = {i\over 2}(b^{-1}-\epsilon)$, $P_0 = i\left({Q\over 2}-\epsilon\right)$, and take $\epsilon\to 0$. Taking the limit, we find
\begin{equation}\label{eq:SvacLimit}\begin{aligned}
	\lim_{\epsilon\to 0}\modS_{P,{i\over 2}(Q-\epsilon)}\left[i({Q\over 2}-\epsilon)\right] &= 2\cosh(2\pi b P)\modS_{P,i{b^{-1}\over 2}}\left[i{Q\over 2}\right] - 2 \modS_{P,{i\over 2}(b^{-1}-b)}\left[i{Q\over 2}\right]\\
	&= 4\sqrt{2}\sinh(2\pi b P)\sinh(2\pi b^{-1}P),
\end{aligned}\end{equation}
precisely reproducing the modular S matrix for the inversion of the Virasoro vacuum character (\ref{eq:modularSVacuumLimit}). Note that one cannot recover this by taking the appropriate limit of (\ref{eq:explicitModularS}), as $\alpha_0 = 2\alpha'$ is at the boundary of the regime of convergence of the integral representation.

\section{Asymptotics of crossing kernels}
In this section we will collect results for the asymptotic form of the elementary crossing kernels when some of the weights are taken to be heavy. These results are important for establishing both the form of our asymptotic formulas and their validity, via the suppression of corrections due to the propagation of non-vacuum primaries.

\subsection{Fusion kernel}

In \cite{Collier:2018exn}, the asymptotic form of the fusion kernel when the S-channel internal weight $P_s$ was taken to be heavy with fixed external weights was extensively studied. The main result of that analysis was the following asymptotic form of the vacuum fusion kernel (\ref{eq:vacFusion}) with pairwise identical operators, which follows directly from the asymptotics of the special function $\Gamma_b$ that were established in that paper
\begin{equation}\begin{aligned}\label{eq:vacFusionAsymptotics}
	\fusion_{P_s\id}\sbmatrix{ P_2 & P_1 \\ P_2 & P_1 } \sim& 2^{-4P_s^2}e^{\pi QP_s}P_s^{4(h_1+h_2)-{3Q^2+1\over 2}}\\
	&\times{2^{Q^2+1\over 6}\Gamma_0(b)^6\Gamma_b(2Q)\over \Gamma_b(Q)^3\Gamma_b(Q+2iP_1)\Gamma_b(Q-2iP_1)\Gamma_b(Q+2iP_2)\Gamma_b(Q-2iP_2)},\,P_s\to\infty
\end{aligned}\end{equation}	
where
\begin{equation}
	\log\Gamma_0(b) = -\int_0^\infty{dt\over t}\left({e^{-Qt/2}\over(1-e^{-bt})(1-e^{-b^{-1}t})}-t^{-2}-{Q^2-2\over 24}e^{-t}\right)
\end{equation}
appears in the large-argument asymptotics of $\Gamma_b(x)$.

By carefully studying the asymptotics of the contour integral in the definition of the fusion kernel, in \cite{Collier:2018exn} it was also established that the fusion kernel with non-zero T-channel weight is exponentially suppressed at large $P_s$ compared to the vacuum kernel 
\begin{equation}\begin{aligned}\label{eq:fusionSuppression}
	{\fusion_{P_sP_t}\sbmatrix{ P_2 & P_1 \\ P_2 & P_1 }\over \fusion_{P_s\id}\sbmatrix{ P_2 & P_1 \\ P_2 & P_1 }}\sim& e^{-2\pi\alpha_t P_s}\left({\Gamma_b(Q+2iP_1)\Gamma_b(Q-2iP_1)\over \Gamma_b({Q\over 2}+i(2P_1-P_t))\Gamma_b({Q\over 2}-i(2P_1+P_t))}\times(P_1\to P_2)\right)\\
	&\times {\Gamma_b(Q-2iP_t)\Gamma_b(-2iP_t)\Gamma_b(Q)^3\over\Gamma_b(2Q)\Gamma_b({Q\over 2}-iP_t)^4},\, P_s\to\infty.
\end{aligned}\end{equation}
Thus we learn that corrections to the heavy-light-light asymptotic formula (\ref{eq:C0HLL}) due to the exchange of non-vacuum primaries in the T-channel are exponentially suppressed.

\subsubsection{With heavy external operators}\label{app:higherGenusAsymptoticCorrections}
In order to establish the validity of the off-diagonal HHL and HHH asymptotic formulas, we need to ensure that the propagation of non-vacuum primaries is suppressed compared to that of the vacuum. The only nontrivial step is establishing the suppression of
\begin{equation}
	{\fusion_{P_2P_2'}\sbmatrix{ P_1 & P_3 \\ P_1 & P_3 }\over \fusion_{P_2\id}\sbmatrix{P_1 & P_3 \\ P_1 & P_3 }}
\end{equation}
when one or both of the external operators $P_1,P_3$ are taken to be heavy along with the S-channel internal weight $P_2$. 

Let's start with the case relevant for the torus two-point kernel. For simplicity and clarity of presentation, we will explicitly present the case where $\alpha_1,\alpha_2 = {Q\over 2}+iP,~P\to\infty$, with $\alpha_3\equiv\alpha_0$ and $\alpha_2'$ fixed. Focusing on the contour integral involved in the definition of the four-point kernel and writing the integration variable as $s = \sigma P$, we have the following asymptotics of the integrand
\begin{equation}
\begin{aligned}
	&\log\prod_{k=1}^4{S_b(s+U_k)\over S_b(s+V_k)}\\
	\sim& \begin{cases} 2\pi(\alpha_0+iQ\sigma)P-\pi i( (Q-\alpha_0)^2+h_2')+\mathcal{O}(P^{-1}),&\text{Im}(\sigma)>2\\
	-2\pi(\alpha_0-iQ\sigma+i\alpha_0\sigma)P-\pi i ( (Q-\alpha_0)^2-h_0+h_2')+\mathcal{O}(P^{-1}),&0<\text{Im}(\sigma)<2\\
	-2\pi(\alpha_0+iQ\sigma)P+\pi i( (Q-\alpha_0)^2+h_2')+\mathcal{O}(P^{-1}),&\text{Im}(\sigma)<0
 	\end{cases}
\end{aligned}
\end{equation}
The integrand decays exponentially at $\sigma = \pm i\infty$ and no poles cross the contour so we evaluate the integral using these leading approximations for the integrand. In this way one finds
\begin{equation}
	\int{ds\over i}\prod_{k=1}^4{S_b(s+U_k)\over S_b(s+V_k)} \sim \text{(order-one)} e^{-2\pi\alpha_0 P},
\end{equation}
so that all together we have
\begin{equation}
	\fusion_{P_2P_2'}\sbmatrix{P_0 & P_1 \\ P_0 & P_1 } \sim \text{(order-one)}(P)^{2h_0-h_2'},
\end{equation}
and corrections due to the propagation of non-vacuum primaries with $0<\alpha_2'<{Q\over 2}$ are encoded by the ratio
\begin{equation}
	{\fusion_{P_2P_2'}\sbmatrix{P_0 & P_1 \\ P_0 & P_1 }\over \fusion_{P_2\id}\sbmatrix{P_0 & P_1 \\ P_0 & P_1 }} \sim \text{(order-one)}P^{-h_2'}.
\end{equation}

The analysis is similar for corrections to the HHH asymptotics due to propagation of non-vacuum primaries in the dumbbell channel. One finds the following for the asymptotics of the integrand when all three weights $\alpha_1,\alpha_2,\alpha_3 = {Q\over 2}+i P$ are taken to be heavy and we scale the integration variable with $P$ as before
\begin{equation}
\begin{aligned}
	&\log \prod_{k=1}^4 {S_b(s+U_k)\over S_b(s+V_k)}\\
	 \sim&\begin{cases}
	3\pi i P^2 + 2\pi i Q\sigma P- {\pi i\over 4}(Q^2+4h_2')+\mathcal{O}(P^{-1}),& \text{Im}(\sigma)>2\\
	-\pi i(1+4i\sigma-\sigma^2)P^2+\pi Q(-2+i\sigma)P-{\pi i\over 12}(-2+Q^2+12h_2')+\mathcal{O}(P^{-1}),& 1<\text{Im}(\sigma)<2\\
	2\pi\sigma P^2 +\pi Q(-2+i\sigma)P-\pi i h_2'+\mathcal{O}(P^{-1}),& 0<\text{Im}(\sigma)<1\\
	\text{c.c.},&\text{Im}(\sigma)<0
	\end{cases}.
\end{aligned}
\end{equation}
In this case the dominant contribution turns out to be of the form
\begin{equation}
	\int{ds\over i}\prod_{k=1}^4{S_b(s+U_k)\over S_b(s+V_k)} \sim \text{(order-one)}e^{-2\pi Q P},
\end{equation}
leading to 
\begin{equation}
	\fusion_{P_2P_2'}\sbmatrix{ P_1 & P_3 \\ P_1 & P_3 } \sim \text{(order-one)}\left({27\over 16}\right)^{3P^2}e^{-\pi Q P}P^{-2h_2'+{5Q^2-1\over 6}}.
\end{equation}
Thus non-vacuum corrections to the HHH asymptotic formula are suppressed via the ratio
\begin{equation}
	{\fusion_{P_2P_2'}\sbmatrix{P_1 & P_3 \\ P_1 & P_3}\over\fusion_{P_2\id}\sbmatrix{P_1 & P_3 \\ P_1 & P_3}} \sim \text{(order-one)}P^{-2h_2'}.
\end{equation}

\subsection{Torus one-point}\label{app:torusOnePtAsymptotics}

In order to establish the validity of the heavy-heavy-light and heavy-heavy-heavy universal formulas, we also need to study the asymptotics of the torus one-point kernel in the limit that the internal weight in one of the channels becomes heavy, namely the limit $P\to\infty$. In this limit, the prefactor $Q_b$ reduces to the following
\begin{equation}\begin{aligned}
	\log Q_b(P,P',P_0) \sim & 2\pi(Q-\alpha_0) P + h_0\log(2P) \\
	&+ \log\sqrt{2}{\Gamma_b(Q+2iP')\Gamma_b(Q-2iP')\over S_b({Q\over 2}+iP_0)\Gamma_b({Q\over 2}+i(2P'-P_0))\Gamma_b({Q\over 2}-i(2P'+P_0))}+\mathcal{O}(P^{-1})
\end{aligned}\end{equation}

To study the asymptotics of the contour integral, we start by considering scaling the integration variable with $P$, ie. $\xi = \sigma P$. Then the integrand behaves in the following way at large $P$ depending on the imaginary part of $\sigma$
\begin{equation}\begin{aligned}\label{eq:logTLargeP}
	\log T_b(\sigma P,P,P_0) \sim \begin{cases}2\pi i \sigma(Q-\alpha_0)P +\mathcal{O}(P^{-1}),~&\text{Im}(\sigma)>1\\ -2\pi(Q-\alpha_0)P+\mathcal{O}(P^{-1}),~&-1<\text{Im}(\sigma)<1\\ -2\pi i\sigma(Q-\alpha_0)P+\mathcal{O}(P^{-1}),~&\text{Im}(\sigma)<-1\end{cases}.
\end{aligned}\end{equation}
In this limit, there are poles extending to the left and right at $\text{Im}(\sigma) = \pm 1$ pinching the contour. 

For $\alpha'$ in the discrete range, we cannot evaluate the integral by deforming the contour and summing over residues e.g. in the $\xi$ right half-plane since the integrand does not decay exponentially along the arc at infinity. However, so long as the internal weight $\alpha'$ obeys the condition (\ref{eq:alphaPrimeCondition}), the integral along the contour ${\rm Re}(\xi) = 0$ converges nicely and the integral in this limit can easily be computed by using the asymptotics (\ref{eq:logTLargeP}). When $\alpha'\in (0,{Q\over 2})$, we have
\begin{equation}\begin{aligned}
	\int_C{d\xi \over i}e^{-4\pi \xi P'}T_b(\xi,P,P_0) \approx {{Q\over 2}-iP_0\over2\pi(-2iP')({Q\over 2}+i(2P'-P_0))}e^{-2\pi P({Q\over 2}+i(2P'-P_0))}.
\end{aligned}\end{equation}
Comining with the asymptotics of the prefactor, we recover the following asymptotics
\begin{equation}\begin{aligned}
	\modS_{PP'}[P_0]\approx& \left({{Q\over 2}-iP_0\over \sqrt{2}\pi(-2iP')({Q\over 2}+i(2P'-P_0))}{\Gamma_b(Q+2iP')\Gamma_b(Q-2iP')\over S_b({Q\over 2}+iP_0)\Gamma_b({Q\over 2}+i(2P'-P_0))\Gamma_b({Q\over 2}-i(2P'+P_0))}\right)\\
	&\times e^{-4\pi iPP'}(2P)^{h_0}
\end{aligned}\end{equation}

To compute the kernel when $\alpha'$ is outside of the regime (\ref{eq:alphaPrimeCondition}), we can make use of the shift relations (\ref{eq:modSShift}).
Note that in the large-$P$ limit, the prefactor on the right-hand side will be exponentially enhanced. So, if $\alpha'+{n\over 2}{\rm Re}(b)>{1\over 2}{\rm Re}(\alpha_0)$ (but $\alpha'+{n-1\over 2}{\rm Re}(b)<{1\over 2}{\rm Re}(\alpha_0))$), then in this limit we have
\begin{equation}
\begin{aligned}
	\modS_{PP'}[P_0] \approx \left(\prod_{k=1}^{n}f(P'-ik{b\over 2},P_0)\right)e^{2\pi n b P}\modS_{P,P'-i{n\over 2}b}[P_0],
\end{aligned}
\end{equation}
where 
\begin{equation}
\begin{aligned}
	f(P',P_0) &= {\Gamma(b({Q\over 2}-i(2P'+P_0)))\Gamma(b({Q\over 2} +i(-2P'+P_0)))\over\Gamma(b(Q-2iP'))\Gamma(-2ibP')}.
\end{aligned}
\end{equation}
Notice that the exponential part of the prefactor cancels the different exponential asymptotics of the shifted kernel $\modS_{P,P'-i{n\over 2}b}$ so that the overall asymptotics are preserved.

\bibliographystyle{JHEP}
\bibliography{universalFormula}
\end{document}